# The Signal Synchronization Function of Myelin

**Authors**: Zhuonan Yu[#1], Peijun Qin[#2], Ruibing Sun[1], Sara Khademi[1], Zhen Xu[1], Qinchao Sun[1], Yanlong Tai[1], Bing Song*[1], Tianruo Guo*[2], Hao Wang*[1]

**Affiliations:**

[1]Institute of Biomedical & Health Engineering, Shenzhen Institute of Advanced Technology (SIAT), Chinese Academy of Sciences (CAS), Shenzhen 518035, China

[2]Graduate School of Biomedical Engineering, University of New South Wales, Sydney, NSW2052, Australia

[#]These authors contributed to the work equally and should be regarded as co-first authors

*These authors are corresponding authors:

***Hao Wang** hao.wang@siat.ac.cn

***Tianruo Guo** t.guo@unsw.edu.au

***Bing Song** bing.song@siat.ac.cn

**Abstract**

The myelinated axons widely present in both central and peripheral nervous systems[1,2]. Its unique compact spiraling structure poses significant challenges for understanding its biological functions and developmental mechanisms[3]. Conventionally, myelin is considered as in insulating layer to achieve saltatory conduction for the enhancement of the neural signal speed[4], which serves as the foundation of neuroscience[5]. However, this insulating hypothesis is inadequate to account for a various experimental observations[6-11], especially the long unmyelinated tract observed in cortex[12,13]. We here show non-random distributions in three ultrastructural features of myelin: the non-random spiraling directions, the localization preferences of myelin's outer tongues, and the radial components along boundaries between oppositely spiraled myelin sheaths. These phenomena are predicted by a novel concept of myelin's biological function, which we propose as the "signal synchronization function." Our findings demonstrate that cytoplasmic channels within myelin may act as coiled inductors, facilitating electromagnetic induction between adjacent myelin sheaths, thereby promoting signal synchronization between axons. This, in turn, explains the non-random ultrastructural features observed. We believe these insights lay the foundation for a new understanding of myelin's inductive function. This perspective could significantly reshape the textbook understanding of myelin and the nervous system by introducing several key points:1. myelin may function as a micro-coil, engaging magnetic fields in neural signaling; 2. axons may not transmit neural signals independently, but instead form clusters to synchronize signal transmission;3. this work provides a clear principle for how electric and magnetic fields influence myelin development, potentially revolutionizing research into myelin regeneration and rehabilitation through electrical and magnetic modulation.

Recently, we proposed an alternative theory to explain the biological functions of myelin, called the Myelin Inductance Theory (MIT)[14-16], whose core hypothesis is that the cytoplasmic channels[17-19] within myelin can form coil inductors, thereby generating magnetic fields (Fig. 1A). This theory provides a concise explanation for many experimental observations, including the mechanisms behind the formation of myelin ultrastructures, such as the same quadrant phenomenon[15], g-ratio[16], and radial sorting[16], as well as how myelinated axons interact with magnetic fields[14], including the mechanisms of magnetic neural stimulation. In this study, we further derive a signal synchronization function (SSF) of myelin, suggesting that neural signals do not propagate independently between axons but instead form internal clusters of synchronized signals (Fig. 1B). This signal synchronization function can influence myelin development, leaving discernible traces on three types of ultrastructural features of myelin observed in the TEM image of the cross-sections of neural bundles:

a. Adjacent myelin sheaths within the same cluster tend to spiral in the same direction, resulting in a frequently observed pattern that all myelin sheaths exhibit the same spiraling direction in a region, as shown in Fig. 1C-i. The black arrows indicate the inner tongues (ITs) and outer tongues (OTs).
b. In regions where myelin sheaths spiral oppositely, a spiral boundary forms, and the outer tongues (OTs) tend to localize along this boundary. As shown in Fig. 1C (ii&iii), the OT of the largest green myelin sheath is situated precisely at the boundary between green (from IT to OT in a clockwise) and red (from IT to OT in a counter-clockwise).
c. The radial components (RCs)[20], also referred to as tight junctions[21-22], tend to localize at the spiral boundary. In TEM images, RCs appear as straight lines with distinct contrast across the compact myelin layers. As observed in Fig. 1C-ii&iii (white arrows in Fig. 1C-ii and purple lines in Fig. 1C-iii), the RCs of the largest green myelin appears along the spiral boundary.

The TEM image (Fig. 1C-ii) will be transformed into a spiral map (Fig. 1C-iii) with indication of ITs, OTs and RCs to enhance illustration clarity, then converted to a polygonal map (Fig. 1C-iv) ensuring that each neighboring polygon pair shares exactly one adjacent edge to enable data processing and modeling, which is to assign random color to each polygon. Moreover, we measure the length of all three boundary types (spiral boundary, non-spiral boundary, and other boundaries not facing or adjacent to any other myelin sheaths) for each myelin sheath, as shown in Fig. 1E, providing the data necessary for the modeling study.

**Experimental Results**

A representative ultra-large TEM image of the mouse optic nerve is shown in Fig. 1D-i, consisting of a 12×12 array of high-resolution TEM images seamlessly stitched together. The corresponding spiral map, derived from this TEM image, is presented in Fig. 1D-ii. It is emphasized that determining the spiraling directions requires only knowing of the leading direction of either the IT or the OT. If neither the IT nor the OT is observable, the spiraling direction cannot be determined, as indicated by the gray areas in the image. The converted polygonal map in Fig. 1D-iii clearly illustrates that polygons of the same color tend to form large, connected regions, suggesting a non-random spiraling pattern. Three TEM samples, each of the same size as Fig. 1D-i, were acquired for analysis. A summary of all measured and modeled data is provided in Table 1. Detailed visualizations and data analysis for each TEM image can be found in **Supplementary S1**. The detailed considerations for the categorization of OTs and RCs to either spiral and non-spiral boundaries is illustrated in **Supplementary S2**. Here we mainly introduce the comprehensive results with inclusion of the data of all three samples.

## 1. The analysis for non-random spiraling

As shown in Table 1-1, the total numbers of spiral ($N_{SB}$) and non-spiral ($N_{NSB}$) boundaries are 303 and 495, leading to a ratio of spiral boundaries ($R_{SB} = N_{SB}/(N_{SB} + N_{NSB})$) of 37.97%, considerably lower than the expected ratio from modeling, which is 50%. A comparison of the ratio of spiral boundaries between the measured and expected value for three samples is in Fig. 1F. The probability density distribution of spiral boundary counts, based on $10^6$ times of modeling, is shown in Fig. 1G-i. The measured value of 303 never occurred, indicated as the red dash line, leading to $P \leq 10^{-6}$. It means, mathematically, the spiraling direction of the myelin sheaths cannot be random.

## 2. The analysis for the non-random distribution of OTs

As presented in Table 1-2, the total number of discernible OTs along the spiral boundaries ($N_{OT-SB}$) and the total OTs ($N_{OT}$) for the three samples are 120 and 281, respectively. Accordingly, the ratio of OTs located on spiral boundaries ($R_{OT-SB} = N_{OT-SB}/N_{OT}$) is 42.7%.

For the modeling, we calculate the ratio of the spiral boundary length to the perimeter for each axon ($\eta_{OT-SB}$), which is explained in Fig. 1E. This ratio represents the probability that the OT of a myelin sheath is located at the spiral boundary, assuming a random distribution. Consequently, it also reflects the expected number of OTs at the spiral boundaries of each myelin sheath. The sum of these ratios ($\sum \eta_{OT-SB}$) provides the total expected number of OTs on spiral boundaries ($N_{OT-SB-R}$) under the assumption of random distribution, which 81.66 in total. Then the expected ratio of OT on spiral boundaries ($R_{OT-SB-R} = N_{OT-SB-R}/N_{OT}$) is 29.06%, which is significant lower than the measured value ($R_{OT-SB}$ =42.7%). A comparison of the ratio of OTs on spiral boundaries between the measured and expected value for three samples is in Fig. 1F.

Since the $\eta_{OT-SB}$ of each myelin sheaths were measured, we also can model the probability density distribution with Monte Carlo method, whose result based on $10^6$ times of modeling is shown in Fig. 1G-ii. The event of the measured value ($N_{OT-SB}$ =120) only occurred once, meaning $P \leq 10^{-6}$. So mathematically, the distribution of OTs is not random. They tend to localize along the spiral boundaries.

## 3. The analysis for the non-random distribution of RCs

As shown in Table 1-3, the total number of discernible RCs along the spiral boundaries ($N_{RC-SB}$), non-spiral boundaries ($N_{RC-NSB}$), and the total RCs ($N_{RC-ALL}$)) are 51, 52 and 140, respectively. Accordingly, the ratio of RCs located on spiral ($R_{RC-SB} = N_{RC-SB}/N_{RC-ALL}$) and non-spiral ($R_{RC-NSB} = N_{RC-NSB}/N_{RC-ALL}$) boundaries is 36.43% and 37.86%.

Since RCs can grow on any myelin sheath, regardless of whether the direction of its spiral can be differentiated, the probability of an RC growing along a spiral boundary should be proportional to the ratio of the total length of spiral boundary ($L_{SB}$) to the total perimeter of all myelin sheaths ($L_{ALL}$). The same principle applies to the non-spiral boundary group. As shown in Table 1-3, the total lengths of the spiral boundary ($L_{SB}$), non-spiral boundary ($L_{NSB}$), and perimeter ($L_{ALL}$) for all myelin sheaths are 8291, 13874 and 46733, respectively (measured in ImageJ without units). Accordingly, the expected ratios of RCs along spiral ($R_{RC-SB-R} = L_{SB}/L_{ALL}$) and non-spiral boundaries ($R_{RC-NSB-R} = L_{NSB}/L_{ALL}$) are 17.74% and 29.69%, respectively. A comparison of the ratio of RCs on spiral boundaries between the measured and expected value is in Fig. 1F. As seen, the measured ratio $R_{RC-SB}$ is more than twice the expected ratio $R_{RC-NSB-R}$ (36.43% / 17.74% = 205.35%), confirming that RCs tend to localize along spiral boundaries. Notably, the measured ratio of RCs on non-spiral boundaries, $R_{RC-NSB}$, also slightly exceeds the expected ratio $R_{RC-NSB-R}$ (37.86% / 29.69% = 127.52%). This result aligns with our theoretical predictions, which will be elaborated upon in the theory section.

Based on the total number of counted RCs ($N_{RC-ALL}$=140) and the expected ratio of RCs on spiral boundaries (17.74%), we applied the Monte Carlo method, running $10^6$ simulations to obtain the probability density distribution, as shown in Fig. 1G-iii. The observed value of 51, indicated by the red line, occurred three times in $10^6$ simulations, implying *P≤3×10⁻⁶*. This suggests that the RCs are statistically inclined to localize along the spiral boundaries.

In summary, the experimental data and modeling results align with the predicted discernible traces on three types of ultrastructural features of myelin, supporting the existence of SSF.

**Myelin Inductance Theory**

In this section, we will mainly introduce the content of the myelin inductance theory (MIT) and explain how it predicts the discernible traces on three types of ultrastructural features of myelin.

1. **The non-random spiraling**

Previously, two types of non-random spiraling in myelin have been reported. The first is the opposite spiraling directions between adjacent myelin sheaths on the same axon[23-24] (Fig. 2A-i), and the second is the same spiraling direction between adjacent myelin sheaths on neighboring axons[25] (Fig. 2A-ii), as also observed in this study. However, previous studies have only observed non-random spiraling in a limited number of cases, preventing any definitive conclusions. MIT provides a comprehensive explanation for these two types of non-random spiraling and predicts that both spiraling types occur simultaneously, inevitably leading to the alignment of the nodes of Ranvier shown in Fig. 2A-iii. The function and mechanism behind its formation is elaborated on as follows.

From a biological functional perspective, the non-random spiraling of myelin sheaths is to ensure that the mutual inductance coefficient between two adjacent myelin sheaths is positive, as explained in Fig. 2B. Assuming that the inductance formed by the cytoplasmic channels within the myelin sheath can generate a magnetic field, the relationship between the internal current direction and the magnetic field direction can be determined by the right-hand rule (Fig. 2B-i). First, consider the case of two adjacent myelin sheaths on the same axon (Fig. 2B-ii). At a certain moment, if the current in myelin *A* flows from the extracellular to the intracellular, and its amplitude is increasing, its voltage will be extracellular positive and intracellular negative. The spiral current generated by this magnetic field can induce a magnetic field and a current in myelin *B*. If we assume this magnetic field is part of the neural signal and its function is also to transmit the neural signal, then this magnetic field should transmit the voltage polarity of myelin *A*, i.e., extracellular positive and intracellular negative, to myelin *B*. According to Lenz's Law, the induced magnetic field in myelin *B* should always oppose the change in myelin *A*. Therefore, based on Lenz's Law and the right-hand rule, we find that only when myelin *A* and myelin *B* have opposite spiraling directions will they have the same voltage polarity. This is referred to in physics as having a positive mutual inductance coefficient. This principle can also be applied to adjacent myelin sheaths on two adjacent axons (Fig. 2B-iii). Only when myelin *A* and myelin *B* have the same spiraling direction will they have the same voltage polarity, i.e., extracellular positive and intracellular negative. According to MIT, these two non-random spiraling phenomena shall happen simultaneously. Therefore, it is inevitably that the myelin sheaths shall have an array pattern as shown in Fig. 2A-iii, with aligned nodes of Ranvier. It is also a unique experimental prediction, whose validation is not included in this study. But we still consider this array pattern as a basic hypothesis, which will be used in the later sections.

From this, it can be seen that the biological function of the non-random spiraling of myelin sheaths is to allow axons to transmit neural signals through electromagnetic induction. This magnetic field will have different effects on the longitudinal direction transmission (i.e., propagation along the axon) and the transverse direction transmission (i.e., propagation across axons) of the neural signal. Our previous research provided detailed theoretical and simulation explanations for the effect of this magnetic field in the longitudinal direction[14]. In simple terms, a positive mutual inductance coefficient between myelin sheaths can reduce the decay of neural signals, thereby increasing the speed of neural signal conduction. Here, we mainly explain the effect of this magnetic field in the transverse direction, as shown in Fig. 2C.

Consider there are two axons with a slight phase difference between their action potentials (APs), where the AP in axon $A$ is slightly ahead of the AP in axon $B$ (Stage 1). Due to the positive mutual inductance between the myelin sheaths, the AP in axon $A$ can induce an identical waveform potential in axon $B$. Thus, the actual potential generated at the same location on axon $B$ is the sum of axon $B$'s AP and the induced potential from axon $A$, which will advance the phase of the AP of axon B a bit. Therefore, when the AP reaches the next section of the axons, the AP on axon $B$ will be excited slightly earlier (Stage 2). Consequently, the phase difference between the APs on axon $A$ and axon $B$ will decrease. This process will continue as the APs propagate along the axons, ultimately resulting in a phase-lock between the APs on axon $A$ and axon $B$, forming synchronized signals (Stage 3). Therefore, one of the biological functions of the non-random spiraling of myelin sheaths is to achieve signal synchronization function (SSF) between axons through electromagnetic induction. In discussion section, we will explain how SSF can enhance the neural signal conduction speed.

From a formation mechanism perspective, the spiraling directions of myelin sheaths are modulated by the magnetic field generated by adjacent myelin sheaths. The detailed mechanism is illustrated in Fig. 2D-F. Consider two adjacent myelin sheaths on two adjacent axons (Fig. 2D&E). The myelin on axon $A$ has formed its spiraling direction, which is anti-clockwise from the outside to the inside (Fig. 2E-i). The myelin on axon $B$ only finishes the 1st layer wrapping, with its two terminals meet each other (Fig. 2E-ii). This is the critical moment for myelin $B$ to form its spiraling direction. Since it is known that the inner terminal is the growing terminal, one of these two terminals shall grow faster to be the inner tongue, and thus, forming the spiraling direction (Fig. 2E-iii).

The electromagnetic induction between myelin $A$ and myelin $B$ is shown in Fig. 2D. When an action potential is generated on axon $A$, there will be a potential, $V_1$, on myelin $A$, and a current, $I_1$, in the cytoplasmic channels of myelin $A$. The current $I_1$ will generate a magnetic field, which is to induces a magnetic field on myelin $B$, generating a potential, $V_2$, on myelin $B$ and a current, $I_2$, in the cytoplasmic channels of myelin $B$.

Since the first phase of the action potential, which is the depolarization phase, is an inward current of $Na^+$ from extracellular to intracellular at the node of Ranvier, the increasing of the intracellular potential will induce a current, $I_1$, in the cytoplasmic channel of myelin $A$, whose direction is from intracellular to extracellular, a clockwise current as shown in Fig. 2E-i. Then due to the positive mutual inductance, the current $I_2$ in myelin $B$ is also clockwise, making the terminal 1 positively charged and terminal 2 negatively charged (Fig. 2E-ii). Here, we need to utilize an important conclusion we obtained earlier[15-16], namely the principle of how E-field modulates the myelin growth: externally applied positive and negative electric fields can inhibit and promote myelin growth, respectively (Hypothesis-E). Since terminal 1 itself is positively charged, it experiences an external negative E-field and should grow faster than terminal 2. So, the depolarization phase is a promoting phase for terminal 1. However, the waveform of the action potential has several phases, who will have a complex effect upon the growth of terminal 1 and 2. Here we need to make a detailed analysis of the waveforms of the current and voltage of these two myelin sheaths as shown in Fig. 2F.

For axon $A$, the waveform of the voltage $V_1$ on its myelin sheath is proportional to the action potential (Fig. 2F-i), which is

$$V_1 \propto AP$$

Since the cytoplasmic channels within the myelin $A$ is capacitively coupled with the extra- and intra-cellular, the current, $I_1$, in the cytoplasmic channels (this is the current to generate the magnetic field) should be proportional to the derivative of $V_1$ (Fig. 2F-ii), which is

$$I_1 \propto dV_1/dt$$

Then the induced current, $I_2$, in myelin $B$ is proportional to the derivative of current in myelin $A$ (Fig. 2F-iii), according to Faraday law of electromagnetic induction,

$$I_2 \propto dI_1/dt$$

The potential, $V_2$, of myelin $B$ is proportional to the charge accumulated on the membrane, which is the integral of current, $I_2$, over time (Fig. 2F-iv),

$$V_2 \propto \int I_2 dt$$

As seen, there will be two promoting phases for terminal 1 and terminal 2, respectively. Here we assume there should be a threshold voltage. Only for the portion of the waveform which is higher than the threshold can promote the myelin growth. The promoting phase for terminal 1, as explain above, is related with the depolarization phase, whose duration is T1 (Fig. 2F-i). While the promoting phase of terminal 2 is related with the repolarization phase, whose duration is T2 (Fig. 2F-i). Since T2 is much longer than T1 (Fig. 2F-i), the promoting phase for terminal 2 is much longer than that of terminal 1 (Fig. 2F-iv). Therefore, terminal 2 will get more promotion upon its growth, and grow faster to form the inner tongue, and finally form an anti-clockwise spiraling direction, which the same as myelin $A$.

In this section, we focus on how the same spiraling phenomenon between neighboring axons is formed by magnetic fields. The formation mechanism for the opposite spiraling of adjacent myelin sheaths on the same axon follows a similar process, so we will not elaborate on it here. In summary, we explain the function and formation mechanism of non-random myelin spiraling from the perspective of magnetic fields, as derived from the MIT.

2. **The non-random distribution of OTs and RCs**

The non-random distribution of OTs and RCs comes from the following three reasons:

a. Hypothesis-E[15-16]: Externally applied positive and negative electric fields can inhibit and promote myelin growth, respectively. As a result, both ITs and OTs tend to appear at locations with strong external positive electric fields. The ITs' behavior leads to the "same quadrant" phenomenon, which has been explained in our previous study[15]. On the other hand, OTs are more likely to occur along the spiral boundaries due to the same mechanism.
b. The RCs tend to grow at the locations with high E-field.
c. Along the boundaries of two clusters, where neural signals are synchronized within each, there exists a region with a high E-field. This high E-field region promotes the growth of RCs while inhibiting the growth of OTs, resulting in a higher probability of their occurrence along the spiral boundaries.

We will elaborate on the three reasons mentioned above, one by one.

**a. Hypothesis-E**

Hypothesis-E was derived from the explanation of the "same quadrant" phenomenon in our previous study[15]. The "same quadrant" phenomenon refers to the tendency of ITs in OLs to position slightly over OTs[11] (Fig. 3A-i&ii), a pattern repeatedly validated by multiple studies[26-32], yet never explained by other existing theories or models. In our theoretical model, we analyzed the electric field (E-field) on the cross-section of the myelin sheath by incorporating the cytoplasmic channel as a low-impedance pathway connecting the IT and OT (Fig. 3A-iii&iv). This creates two high-current zones (also high E-field zones) at the locations of the IT and OT (Fig. 3B-i). The interaction between these two high E-field zones leads to periodic changes in the potential upon IT during its circular growth, resulting in peaks with opposite polarities when the IT is slightly before (Position 1) and over (Position 2) the OT (Fig. 3B-ii). By combining this potential curve with position (Fig. 3B-ii) and the "same quadrant" phenomenon (Fig. 3A-ii), we conclude that an external negative (internally positive) E-field promotes IT growth (lowest occurrence in Fig. 3A-ii), while an external positive (internally negative) E-field inhibits IT growth (highest occurrence in Fig. 3A-ii), termed Hypothesis-E. This explanation for the "same quadrant" phenomenon is a significant breakthrough, as it successfully explains the abrupt change in occurrence between Position 1 and Position 2 (Fig. 3A-ii), which has been the most puzzling aspect of this phenomenon. This principle applies not only to ITs but also to OTs. Therefore, we predict that OTs will tend to stay in regions with an externally applied positive E-field, which corresponds to the spiral boundaries that will be discussed in the following section.

**b. RCs tend to grow at locations with high E-field**

It has been repeatedly reported that RCs tend to appear at the positions of ITs and OTs in OLs, and occasionally at the interfaces with adjacent axons (Fig. 3B-iii)[20]. These locations of RCs significantly overlap with the high E-field zones we identified in Fig. 3B-i, suggesting that RCs may preferentially grow in regions of high E-field. Based on this, we further predict that RCs are likely to appear along the spiral boundaries where there are high E-field regions.

**c. The high E-field along spiral boundaries leads to preference of OTs and RCs on spiral boundaries**

Consider the scenario of two clusters with opposite spiraling directions (Fig. 3C-i), where neural signals within each cluster are synchronized. For two adjacent axons within the same cluster, there will be no cross-axon electric field (E-field) since their synchronized spiking-resting states create no potential difference between them (Fig. 3C-ii). However, for adjacent axons located along the boundary between two clusters, an asynchronous spiking-resting state will cause a potential difference, generating a cross-axon E-field (Fig. 3C-iii). As a result, a cross-boundary E-field is generated along the spiral boundaries (Fig. 3C-i), which explains why RCs tend to localize at spiral boundaries (Fig. 3C-iv).

In the scenario shown in Fig. 3C-iii, when axon *B* is spiking, it emits an outward positive E-field since the AP is typically a monophasic positive voltage pulse, with the extracellular as the reference. As a result, the OT of axon *A* experiences an externally applied positive E-field from axon *B* at the interface, inhibiting its growth. This explains why OTs tend to localize along the spiral boundaries (Fig. 3C-iv).

Now, let's consider a scenario where three clusters are adjacent to each other (Fig. 3D-i), a situation that will inevitably occur. If clusters *A* and *B* have opposite spiraling directions, then cluster *C* will inevitably merge with either *A* or *B* in terms of spiraling direction (Fig. 3D-ii). This implies that spiral boundaries do not represent the entirety of the cluster boundaries. It is clear that spiral boundaries account for only half of the total cluster boundaries, while the other half lies within non-spiral boundaries. Therefore, while spiral boundaries are definite cluster boundaries, only a portion of non-spiral boundaries serve as cluster boundaries. This explains why OTs and RCs show a higher preference for localizing along spiral boundaries compared to non-spiral boundaries.

In summary, we have explained how the Myelin Inductance Theory (MIT) predicts the discernible traces on three types of ultrastructural features of myelin.

**Discussion**

1. **The influence of the signal synchronization function (SSF)**

We believe that one of the functions of SSF is to enhance the conduction speed of neural signals, which may also be one of the primary function of myelin, as illustrated in Fig. 4A-C. Consider a bundle of axons with aligned nodes that belong to the same cluster (Fig. 4A-i). The cross-section at the node positions of these four axons in Fig. 4(A-ii) shows a scenario where they share a common extracellular space. As a result, they also share a common extracellular potential ($V_e$), and since the action potential (AP) is defined as the potential difference between the intracellular and extracellular spaces ($AP = V_i - V_e$), this shared $V_e$ would influence the overall conduction dynamics.

If the signals of these four axons are not synchronized, at any given moment, only one of them may be spiking (Fig. 4B-i), activating an inward $Na^+$ current and generating a depolarization phase with a relatively small slope (Fig. 4B-ii). However, if all four axons spike simultaneously, activating their inward $Na^+$ currents together (Fig. 4B-i), the shared extracellular potential ($V_e$) will change much more rapidly, resulting in a steeper depolarization phase (Fig. 4B-ii).

The scenario depicted in Fig. 4B has several important effects. First, the amount of $Na^+$ required for AP activation in each axon is significantly reduced, leading to lower energy consumption. Additionally, a steeper depolarization phase indicates faster spiking dynamics, which translates to an increased conduction speed. A model demonstrating this effect is shown in Fig. 4C. **XXXXXXX The modeling details can be found in supplementary ?.**

A potential issue with this model comes from the consideration in myelin sheath length. It is well-established that the length of myelin sheaths is closely related to axonal diameter, with larger axons tending to have longer sheaths[33]. However, node alignment requires uniform sheath lengths across axons. As a result, node alignment may only occur between axons with similar diameters, as shown in Fig. 4D. In particular, axons with significantly larger diameters, which we occasionally observe in TEM images, may function independently. This suggests that axons of different sizes may have distinct functional roles: smaller axons may cluster together to enhance conduction speed, while larger axons may function more independently.

## 2. A retrospect of MIT

The MIT suggests that the magnetic field generated by channel inductance influences the spiraling directions of myelin, a concept that extends beyond the scope of conventional theories and models. This raises the question: is there an alternative theory that can explain the non-random spiraling while remaining consistent with traditional frameworks? To explore this, we propose a thought experiment, as illustrated in Fig. 4E.

If we assume that the two types of non-random spiraling occur simultaneously, then myelin *A* and *B* should spiral in opposite directions, while myelin *A* and *C* should spiral in the same direction (Fig. 4E-i). Let's further assume that this non-random spiraling is induced by some chemical or biological effect, for example, by a protein released from myelin *A*. In this case, the information carried by this protein from myelin *A* to *B* would indicate "opposite spiraling", while from *A* to *C* it would indicate "same spiraling." This would imply that the protein functions in opposite ways along different pathways, which seems highly improbable.

However, if we assume that this non-random spiraling is induced by a physical field—such as a magnetic field emitted from myelin *A*—then the directions of this magnetic field at myelin *B* and *C* would indeed be opposite (Fig. 4E-ii), consistent with the opposite spiraling between myelin *B* and *C*. Therefore, under the assumption that the two types of non-random spiraling occur simultaneously, an alternative plausible theory from a biological or chemical perspective will be extremely challenging. Of course, whether these two types of non-random spiraling actually occur simultaneously will be a key focus of future research.

**Conclusion**

In this study, we demonstrate the discernible traces of three types of ultrastructural features of myelin predicted by the Myelin Inductance Theory (MIT). Specifically, these features include the non-random spiraling directions of myelin sheaths and the localization preferences of the outer tongues and radial components of myelin sheaths along the spiral boundaries. We also provide a detailed explanation of MIT, whose core hypothesis posits that the cytoplasmic channels within the myelin sheaths form low-impedance pathways, generating magnetic fields during action potential activation, much like a coil inductor. We outline the process by which the three predictions are derived from the theory and conclude that myelin has a signal synchronization function (SSF), which is not accounted for in conventional theories.

**Acknowledgement**

**Author contribution**

**Method**

1. TEM sample preparation protocol

A 3-month-old C57BL/6 mouse was anaesthetized using a combination of Zoletil (55 mg/kg body weight) and Xylazine (13.75 mg/kg body weight). Anesthesia was confirmed by the absence of a withdrawal reflex in response to hind paw pinching. Once deep anesthesia was achieved, transcardiac perfusion was carried out with 30 ml of ice-cold 0.9% saline solution to prevent blood coagulation, indicated by the liver turning pale. This was followed by perfusion with 30 ml of ice-cold fixative solution, comprising 4% paraformaldehyde (PFA) and 1% glutaraldehyde (both Sigma-Aldrich, USA), dissolved in Dulbecco's Phosphate-Buffered Saline (DPBS) (Servicebio, China). After perfusion, the mouse was decapitated, and the scalp was removed to expose the skull. Lateral incisions were made on both sides of the skull, and a cut was made along the sagittal suture from the brainstem. The skull was then removed to reveal the entire brain. Using blunt curved tweezers, the anterior portion of the brain was gently lifted to expose the optic nerves, and a 2 mm section of the optic nerve was excised with Vannas scissors. To facilitate optimal fixation in the subsequent step, superficial incisions were made on the epineurium of the harvested optic nerve segment with scalpel. The tissue was further fixed in DPBS containing 4% PFA and 2.5% glutaraldehyde for 2 hours at room temperature, then transferred to 4°C environment to fix for an additional 24 hours. Following fixation, the optic nerve sample was rinsed three times with 0.1M phosphate buffer (pH 7.0), each rinse lasting 15 minutes, to remove residual paraformaldehyde and glutaraldehyde. The tissue was then post-fixed in 1% osmium tetroxide solution for 2 hours, followed by an additional three rinses with 0.1M phosphate buffer under the same conditions. The sample was dehydrated through a graded ethanol series (30%, 50%, 70%, and 80%), with each step lasting 15 minutes. Subsequently, the sample was transitioned through 90% and 95% acetone for 15 minutes each. This was followed by two treatments with 100% acetone, each lasting 20 minutes. For resin embedding, the tissue was first infiltrated with a 1:1 (v/v) mixture of Spurr embedding resin (Sigma-Aldrich, USA) and acetone for 1 hour, then with a 3:1 (v/v) mixture for 3 hours. The sample was then incubated in pure Spurr embedding resin overnight at room temperature. Polymerisation of the resin was carried out at 70°C overnight, resulting in the permanent embedding of the optic nerve tissue. Ultra-thin cross-sections of the embedded sample were prepared using an EM UC7 Ultramicrotome (Leica Microsystems, Germany). These sections were stained with lead citrate solution and 50% saturated uranyl acetate in ethanol for 5–10 minutes each. Micrographs of the stained sections were captured using an HT7800 transmission electron microscope (Hitachi High-Tech, Japan).

2. TEM image processing method

For each sample, 144 high resolution TEM images (×25000) were recorded as an 12×12 array, each image has 15% area overlapped with neighboring images to enable image stitching. Then these images were stitched by ImageJ with the plugin function (Plugins-Stitching-Grid/Collection stitching) to form a complete ultra-large TEM image. The spiral and polygonal maps were plotted based on the complete TEM images in Microsoft Visio. The length measurement was conducted based on the spiral map in ImageJ.

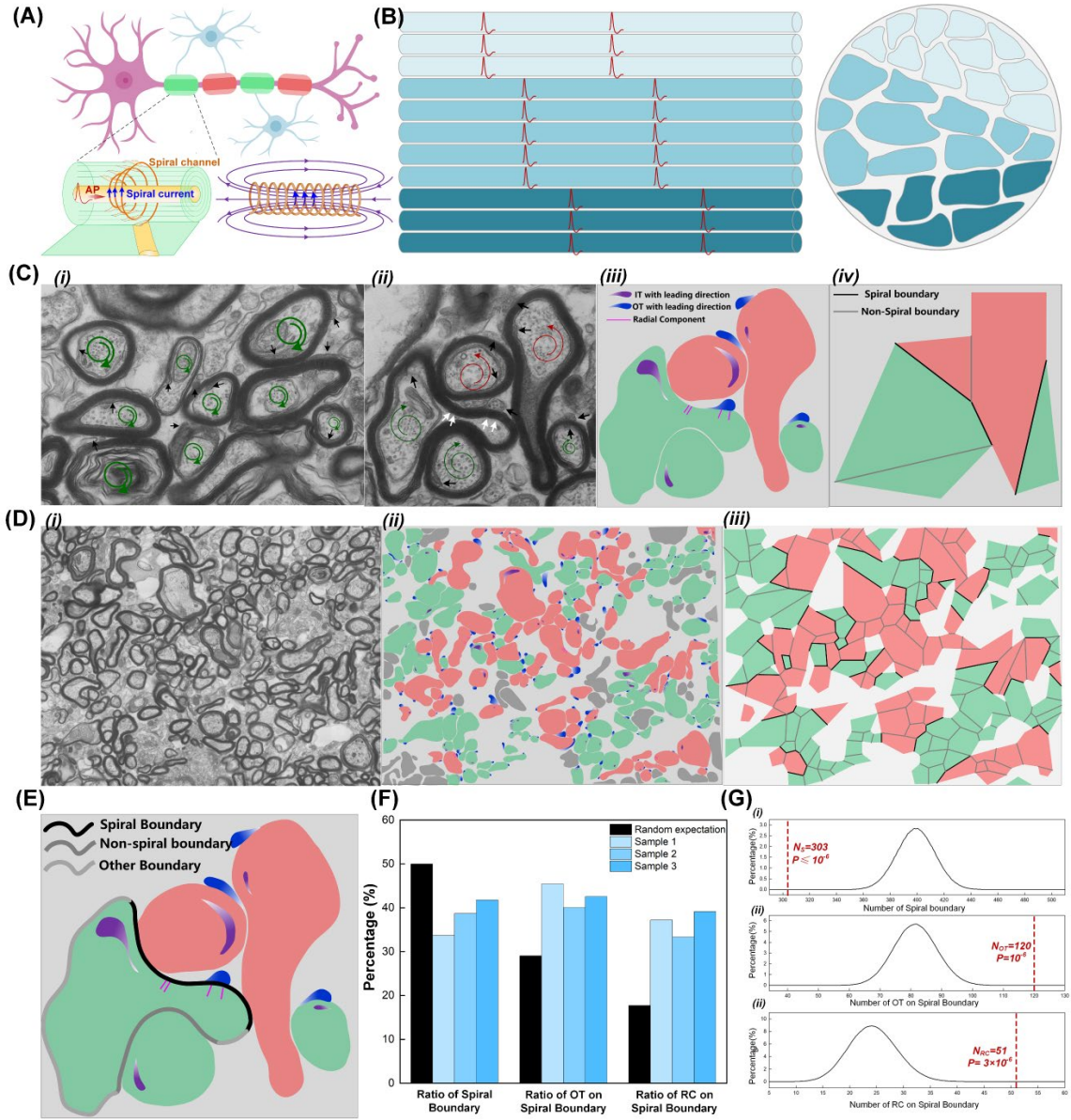

Fig. 1. (A) Myelin Inductance Theory (MIT); (B) The Signal Synchronization Function (SSF); (C): (i) The sample of the same spiraling, the black arrows indicate the IT/OT; (ii) The sample of the area of oppositely spiraling, the black and white arrows indicate the IT/OT and RC, respectively; (iii) The spiral map with indication of IT/OT and RC; (iv) The polygonal map showing the neighboring relationships of (iii); (D): (i) A high-resolution TEM of mouse's optical nerve; (ii) The spiral map of (i); The polygonal map of (ii); (E) The method to calculating the length of spiral boundary, non-spiral boundary and other boundary (the boundary not facing or adjacent to other myelin sheaths); (F) The comparison of the expectation from modeling and the experimental data; (G) The position of the experimental data in the probability density distribution from modeling, indicating the $P$ value.

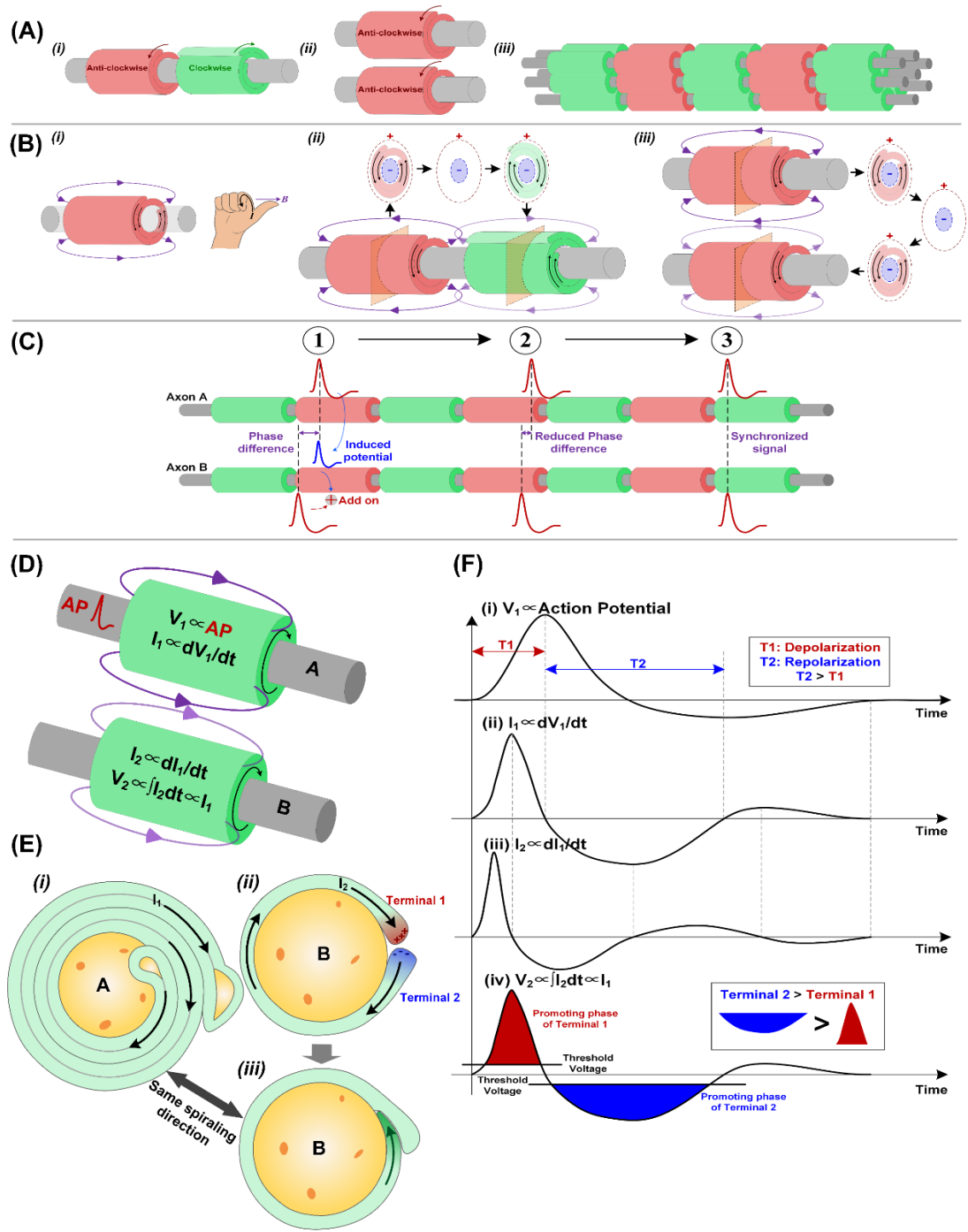

Fig. 2. (A): (i) The opposite spiraling; (ii) The same spiraling; (iii) The alignment of nodes of Ranvier; (B): (i) The magnetic field by the myelin inductance to be determined by right-hand rule; the positive mutual inductance for adjacent myelin sheaths on the same axon (ii) and neighboring axons (iii); (C) The signal synchronization function by the electromagnetic induction; (D) The voltage-current relationship of two adjacent myelin sheaths; (E) The formation mechanism of same spiraling; (F) A detailed waveform analysis of the voltage-current relationship in (D).

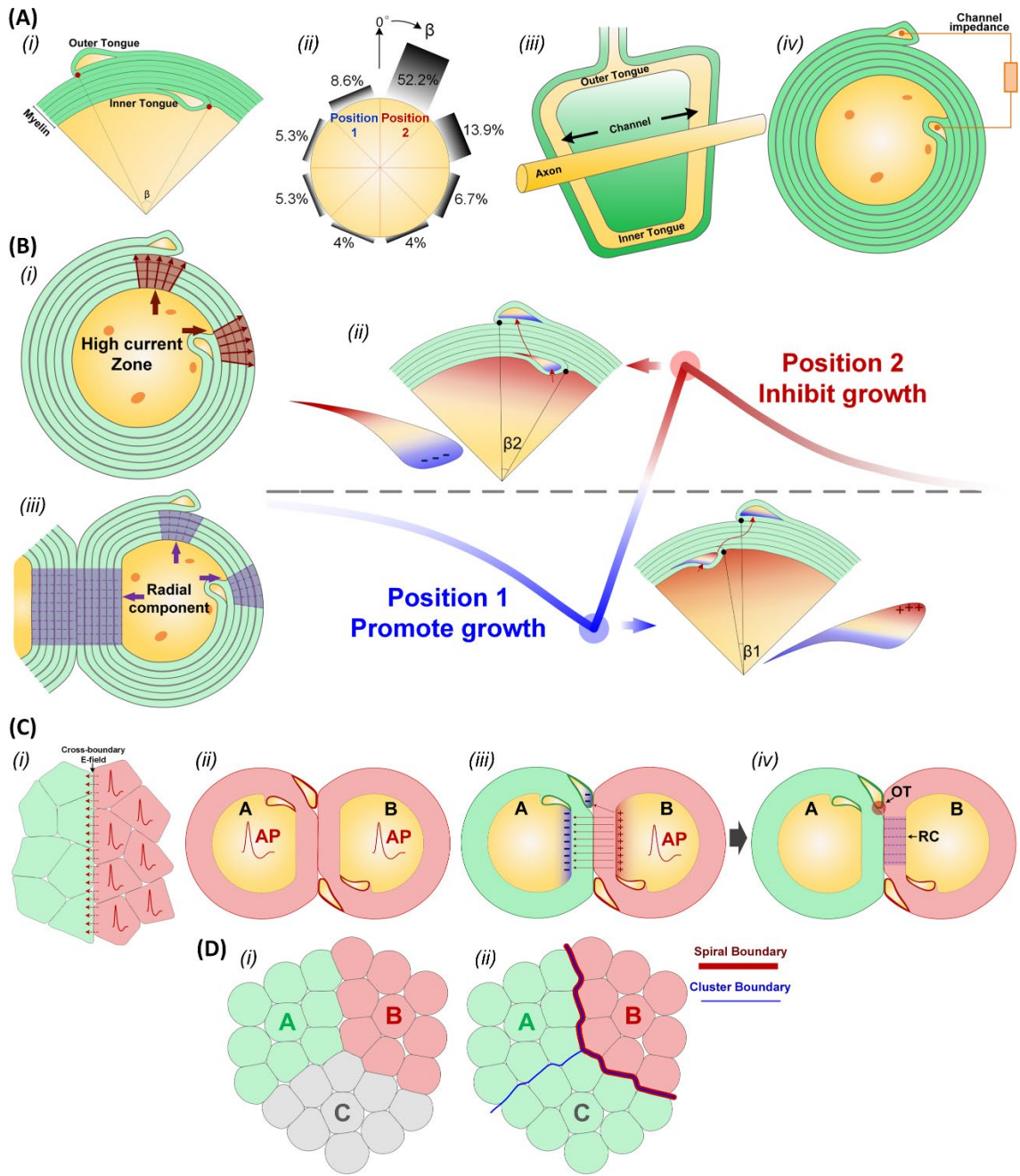

Fig. 3. (A&B) A revisit of the explanation for same quadrant phenomenon; (A): An illustration (i) and a duplication of data (ii)[11] of the same quadrant phenomenon; An illustration of the unwrapped myelin sheaths (iii) which models the cytoplasmic channel as an impedance connecting the IT and OT (iv); (B): (i) The two high-current zones obtained from the modeling; (ii) An illustration of potential waveform upon IT during its circular growth; (iii) The distribution preference of RCs; (C): (i) A cross-boundary E-field at cluster boundaries; (ii-iv) The cross-axon E-field for synchronized and asynchronized situations; (D): The situation of three clusters neighboring to each other (i) and resulted cluster boundaries (ii).

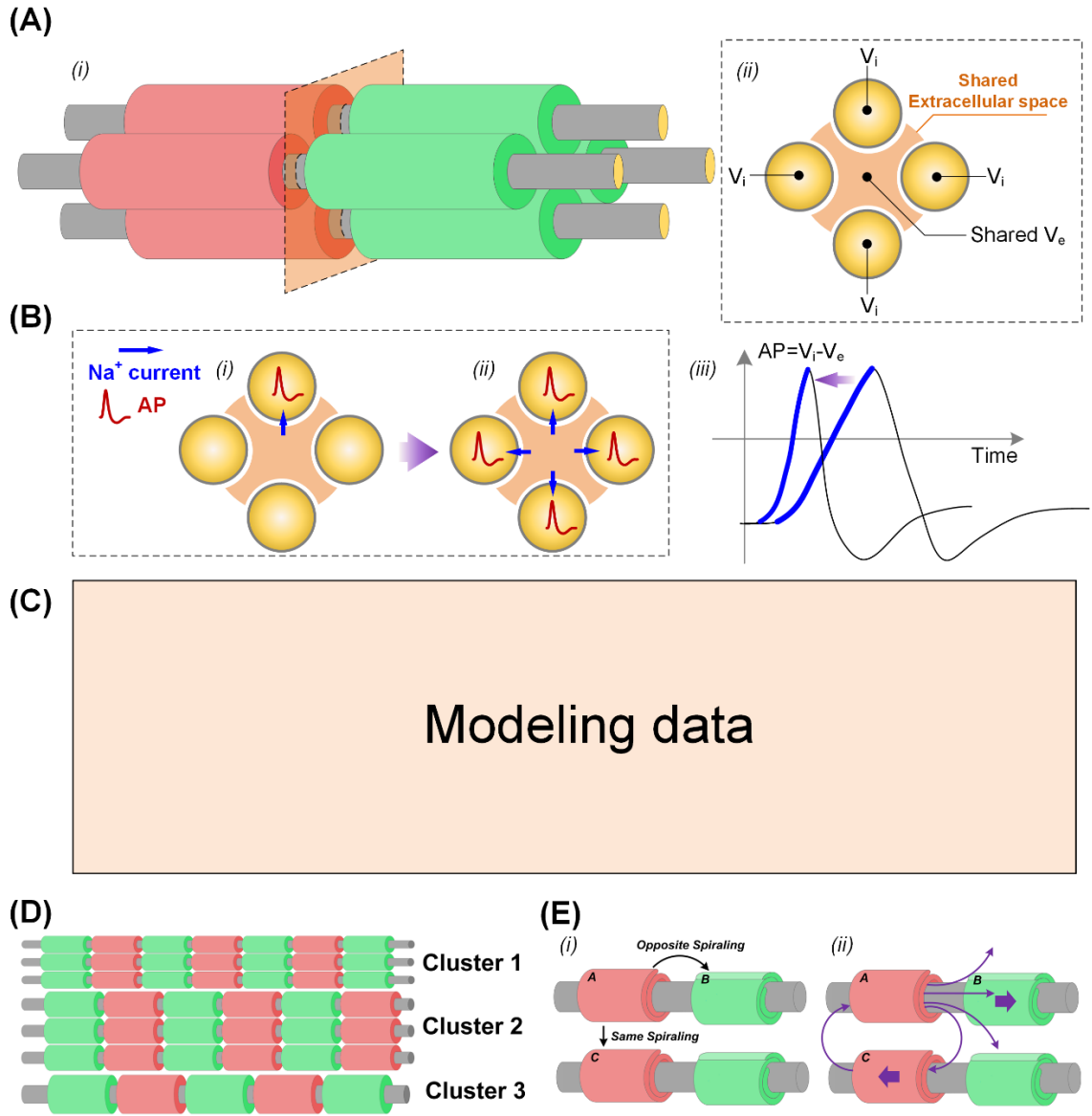

Fig. 4. (A): (i) The scenario of four axons with aligned nodes of Ranvier; (ii) The cross-section of the aligned nodes of Ranvier, showing the shared extracellular space of these aligned nodes; (B): (i) An comparison of the Na$^+$ current for the situations when the APs of these axons are asynchronized and synchronized; (ii) The comparison of the AP waveforms for the two situations in (i); showing that synchronized APs can induce a steeper slop of the depolarization phase; (C) The modeling results for the influence of the SSF:XXX; (D) An illustration of the situation of nodes alignment for axons with difference axonal diameters; (E) A thought experiment for possible alternative theories to explain non-random spiraling: (i) The information release from myelin A to myelin B and C is opposite; (ii) The directions of the magnetic field from myelin A to myelin B and C are opposite, aligning with the opposite information release in (i).

**Table 1 The summary of all measurement results**

| Measured results | Sample 1 | Sample 2 | Sample 3 | Total |
|---|---|---|---|---|
| **1. Non-random spiraling** | | | | |
| Number and ratio of Spiral boundaries $N_{SB}$; $R_{SB} = N_{SB}/N_I$ | 95 | 96 | 112 | 303 |
| | 33.69% | 38.71% | 41.79% | 37.97% |
| Number of non-spiral boundaries, $N_{NSB}$ | 187 | 152 | 156 | 495 |
| Number of Internal boundaries, $N_I = N_{SB} + N_{NSB}$ | 282 | 248 | 268 | 798 |
| Expected ratio of spiral boundary with $10^6$ times of modeling, $R_{SB-R}$ | 50% | 50% | 50% | 50% |
| **2. Non-random distribution of OTs** | | | | |
| Number and ratio of discernible OTs on spiral boundaries $N_{OT-SB}$; $R_{OT-SB} = N_{OT-SB}/N_{OT}$ | 40 | 40 | 40 | 120 |
| | 45.5% | 40.4% | 42.55% | 42.7% |
| Number of discernible OTs on Myelin along spiral boundaries, $N_{OT}$ | 88 | 99 | 94 | 281 |
| Random number and ratio of OTs expected on spiral boundaries $N_{OT-SB-R}$; $R_{OT-SB-R} = N_{OT-SB-R}/N_{OT}$ | 25.8 | 28.66 | 27.2 | 81.66 |
| | 29.32% | 28.95% | 28.94% | 29.06% |
| **3. Non-random distribution of RCs** | | | | |
| Number and ratio of discernible RCs on spiral boundaries, $N_{RC-SB}$; $R_{RC-SB} = N_{RC-SB}/N_{RC-ALL}$ | 16 | 17 | 18 | 51 |
| | 37.21% | 33.33% | 39.13% | 36.43% |
| Number and ratio of discernible RCs on non-spiral boundaries $N_{RC-NSB}$; $R_{RC-NSB} = N_{RC-NSB}/N_{RC-ALL}$ | 16 | 23 | 14 | 52 |
| | 37.21% | 45.1% | 30.43% | 37.86% |
| Number of discernible RCs on all boundaries $N_{RC-ALL}$ | 43 | 51 | 46 | 140 |
| Total spiral boundary length of all myelin and expected ratio of RCs on spiral boundaries $L_{SB}$; $R_{RC-SB-R} = L_{SB}/L_{ALL}$ | 2587 | 2804 | 2900 | 8291 |
| | 15.33% | 18.66% | 19.55% | 17.74% |
| Total non-spiral boundary length of all myelin and expected ratio of RCs on non-spiral boundaries, $L_{NSB}$; $R_{RC-NSB-R} = L_{NSB}/L_{ALL}$ | 5199 | 4544 | 4131 | 13874 |
| | 30.82% | 30.23% | 27.85% | 29.69% |
| Total perimeter of all myelin, $L_{ALL}$ | 16871 | 15030 | 14832 | 46733 |

# Supplementary Information of
# The Signal Synchronization Function of Myelin

**S1: The detailed image and data of all three sample.**

1. **The details of sample 1**

The detailed images and data of sample 1 are shown from Figure S1.1-S1.6.

   a. **The analysis of the non-random spiraling**

As shown in Figure S1.3, there are 95 spiral boundaries and 187 non-spiral boundaries. Thus, the ratio of spiral boundaries is $95/(95 + 187) = 33.69\%$. The modeling based on the polygonal map shown in Figure S1.3 generates the probability density distribution shown in Figure S1.6(a). The event of 95 spiral boundaries did not occur in $10^6$ times of modeling, indicating $P \leq 10^{-6}$.

   b. **The analysis of OTs along spiral boundaries**

As shown in Figure S1.4, there are 88 axons with discernible OTs along the spiral boundaries. Among them, there are 40 OTs along the spiral boundaries, indicated with purple circles. Thus, the ratio of OTs along spiral boundaries is $40/88 = 45.5\%$. The modeling based on the measurement of the ratio of spiral boundaries for each axons generates the probability density distribution shown in Figure S1.6(b). The event when the number of OTs along spiral boundaries is not lower than 40, which is the measured value, takes 0.0375%, indicate $P \leq 0.000375$.

   c. **The analysis of RCs along spiral boundaries**

As shown in Figure S1.5, there are 43 groups of discernible RCs, while 16 of them are along the spiral boundaries, indicated with purple circles. Thus, the ratio of RCs along spiral boundaries is $16/43 = 37.21\%$. Meanwhile, the number of RCs along non-spiral boundaries is 16, leading to the ratio of RCs along non-spiral boundaries is $16/43 = 37.21\%$.

The total length of perimeter of all axons is 16871, while the total length of spiral boundaries and non-spiral boundaries are 2587 and 5199, respectively. Thus, the expected ratio of RCs along spiral and non-spiral boundaries should be $2587/16871 = 15.33\%$ and $5199/16871 = 30.82\%$ under the assumption of random distribution. This the measured value of ratio of RCs along spiral boundaries is more than twice of the expected value ($37.21\%/15.33\% = 242.73\%$), while the measured value of ratio of RCs along non-spiral boundaries is slightly higher than the expected value ($37.21\%/30.82\% = 120.73\%$).

The modeling based on the expected ratio (15.33%) and the total number of RCs (43) generates the probability density distribution shown in Figure S1.6(c). The event when the number of RCs along spiral boundaries is not lower than 16, which is the measured value, takes 0.0392%, indicate $P \leq 0.000392$.

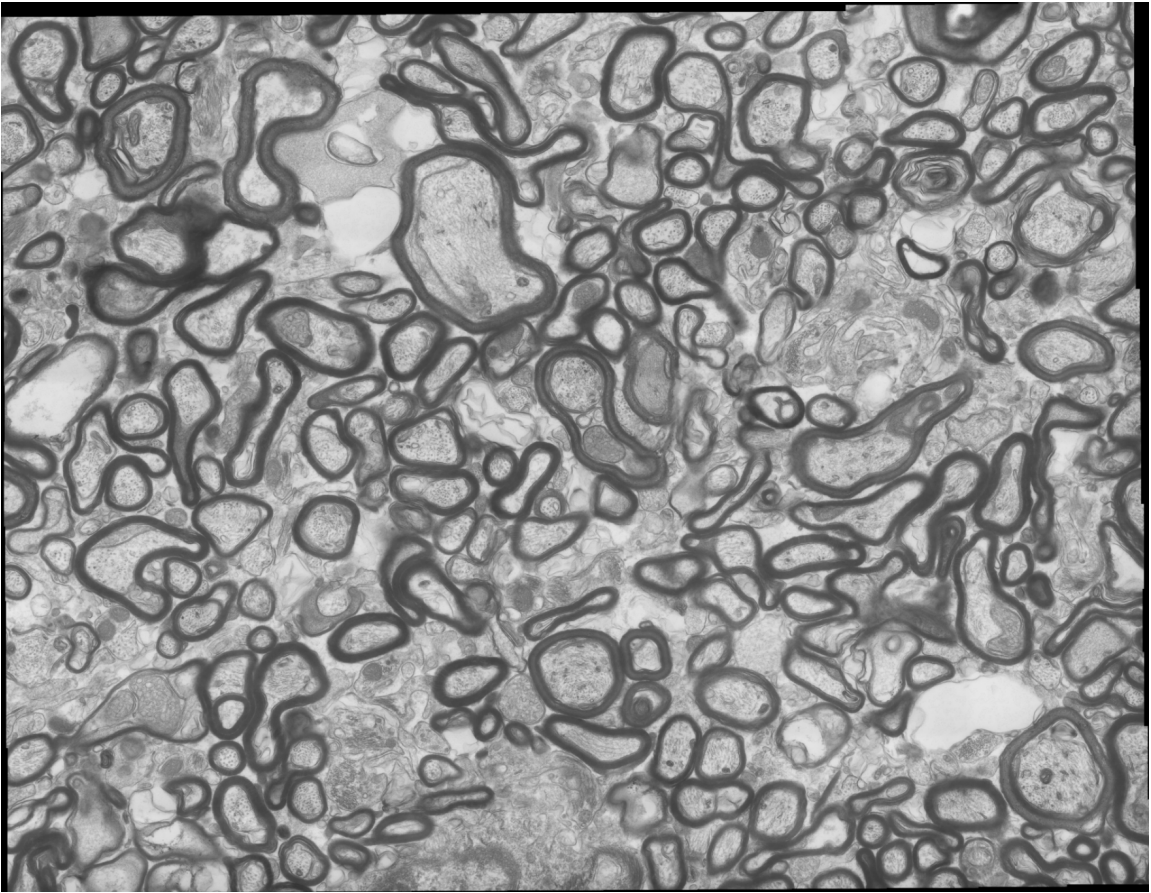

Figure S1.1 The ultra-large TEM image of sample 1 showing the optical nerve of mouse

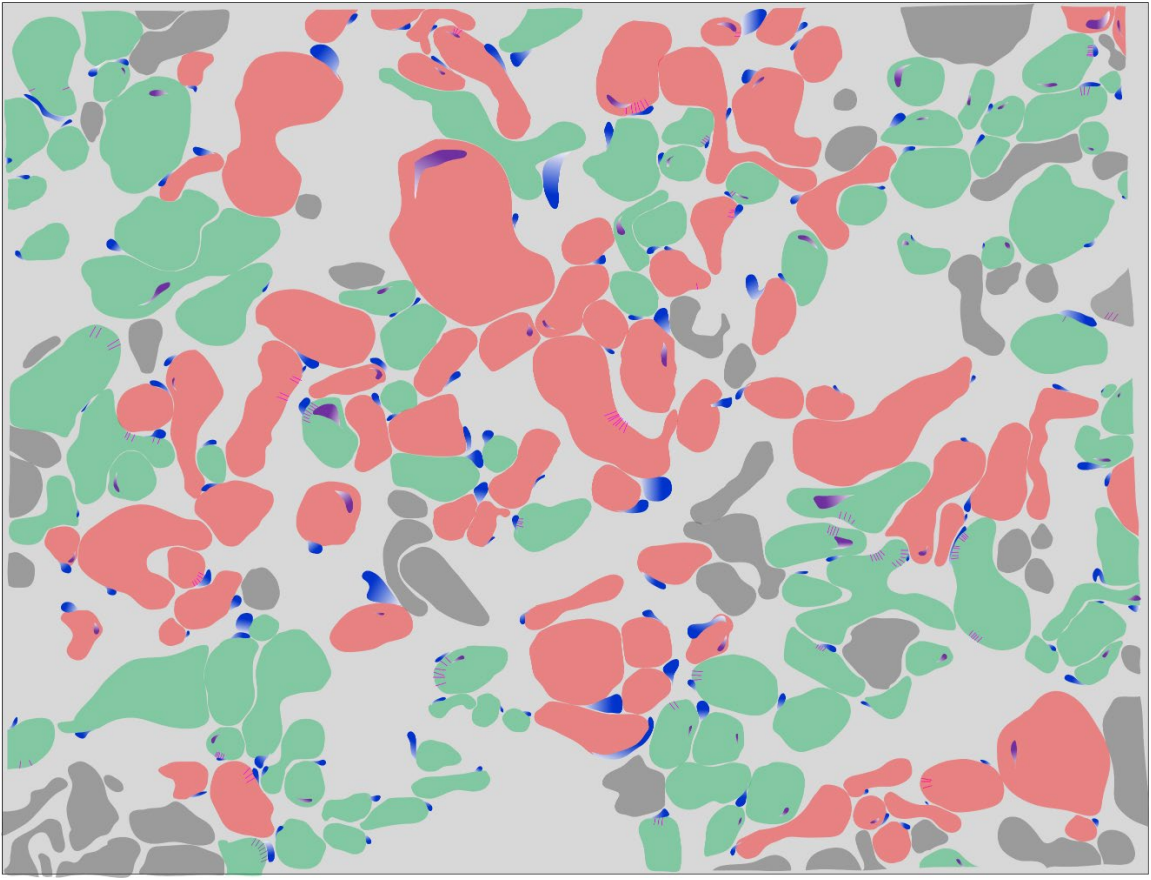

Figure S1.2 The spiral map of the TEM image of sample 1

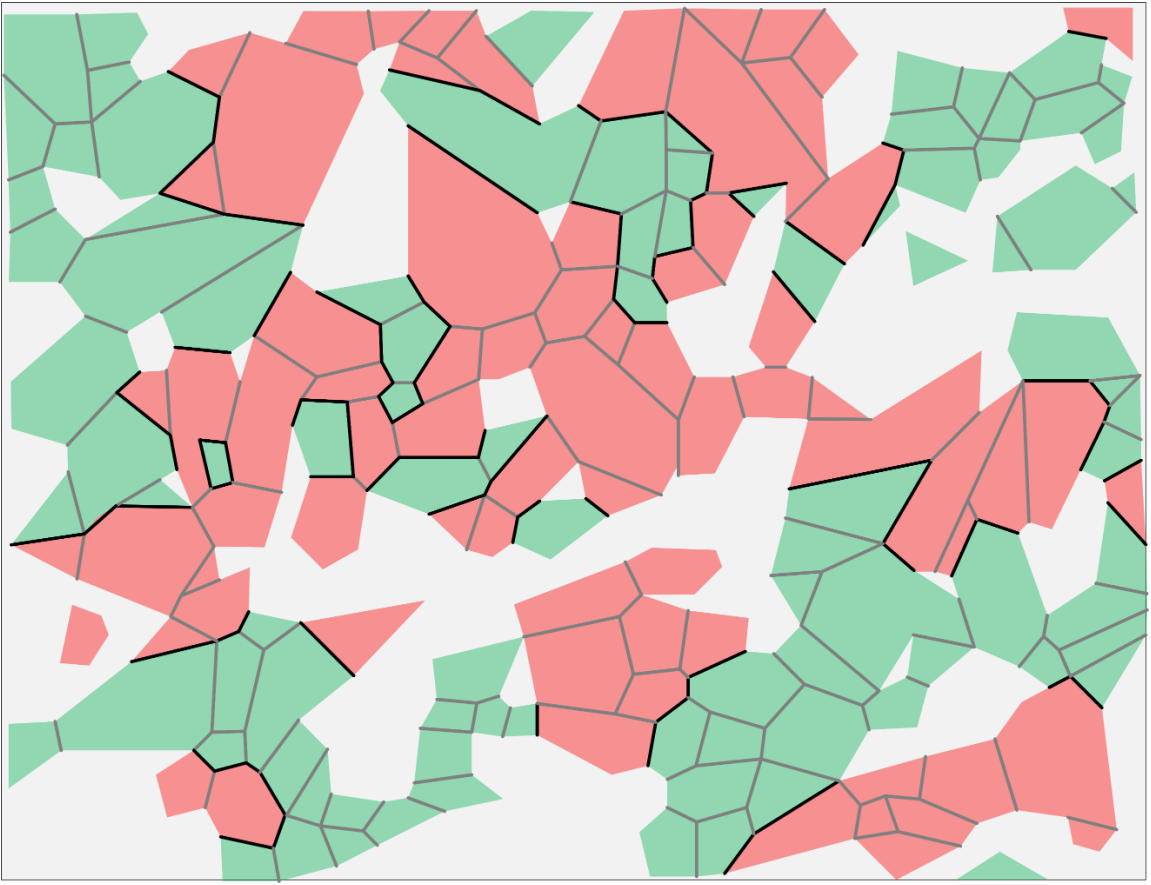

Figure S1.3 The polygonal map of sample 1

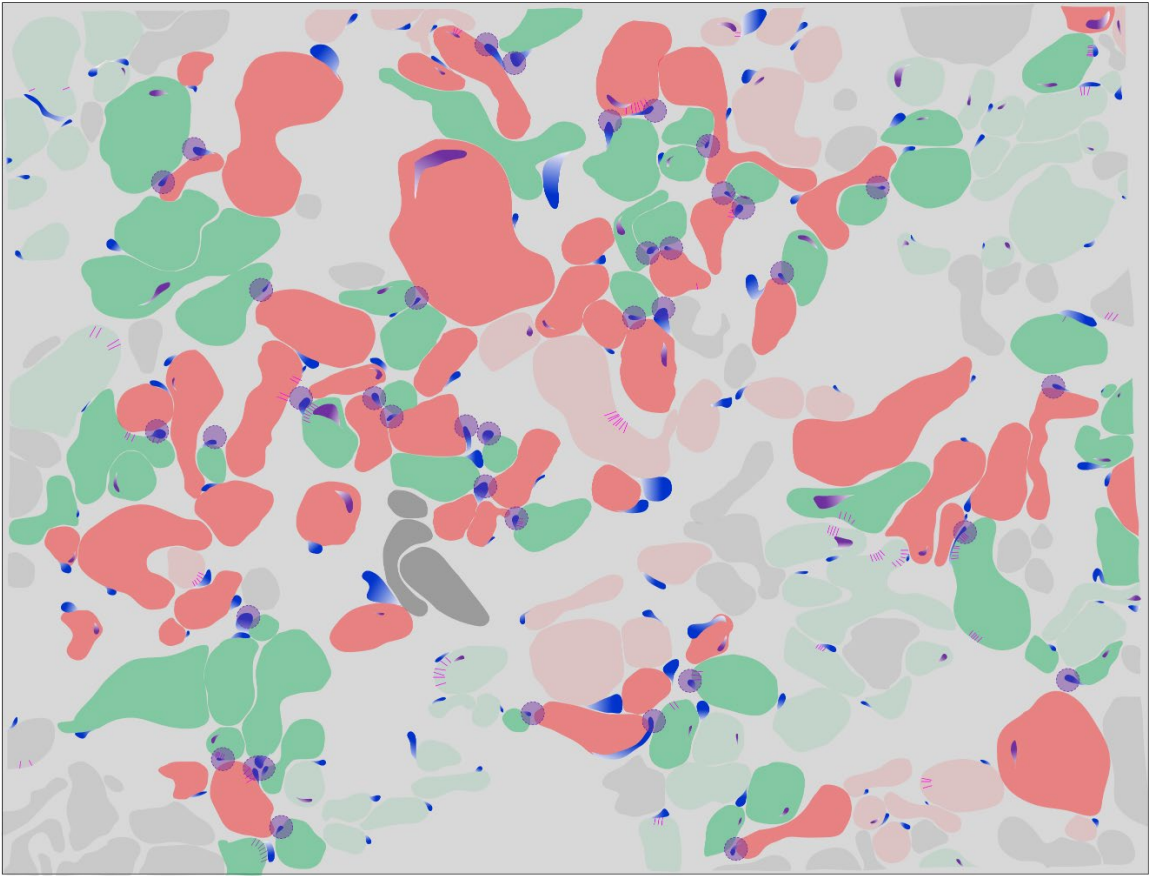

Figure S1.4 The axons along the spiral boundaries, with all 40 OTs at the spiral boundaries indicated with purple circles

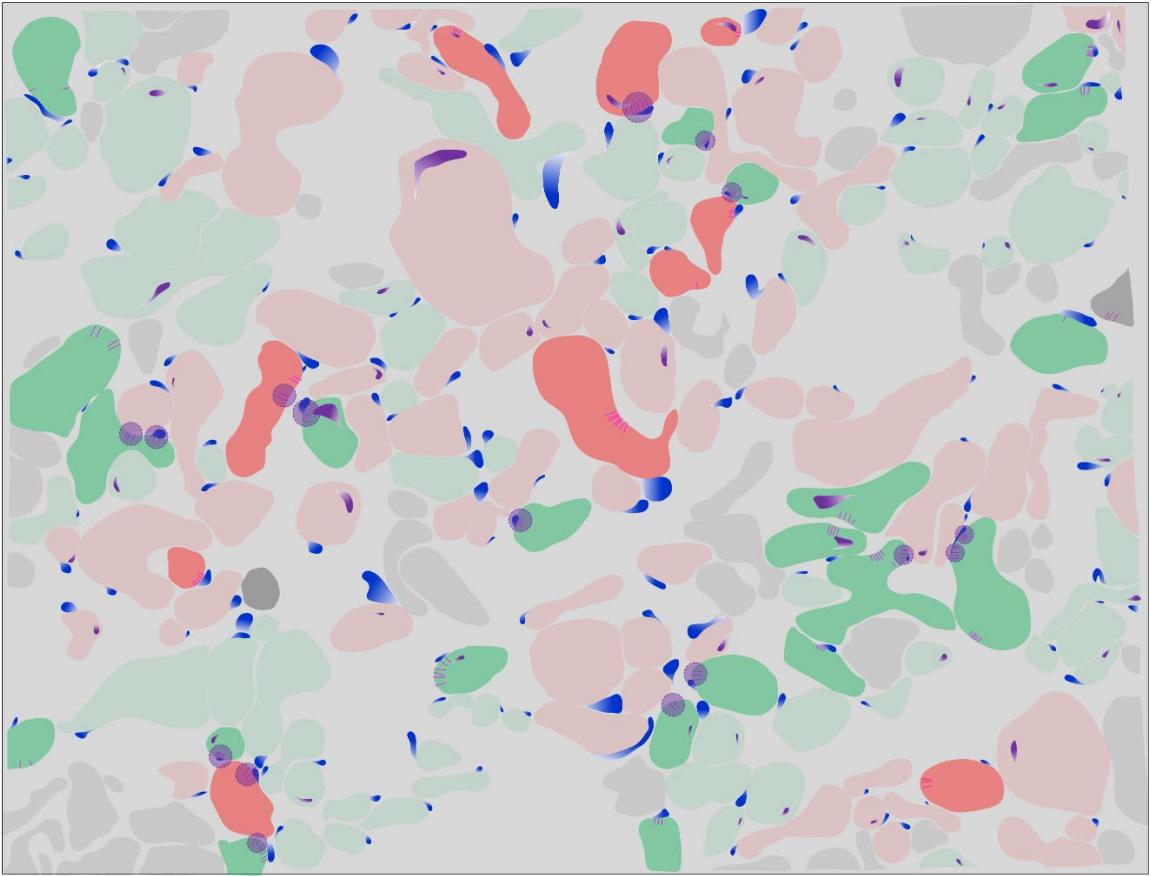

Figure S1.5 The axons with RCs, with all 16 RCs at the spiral boundaries indicated with purple circles

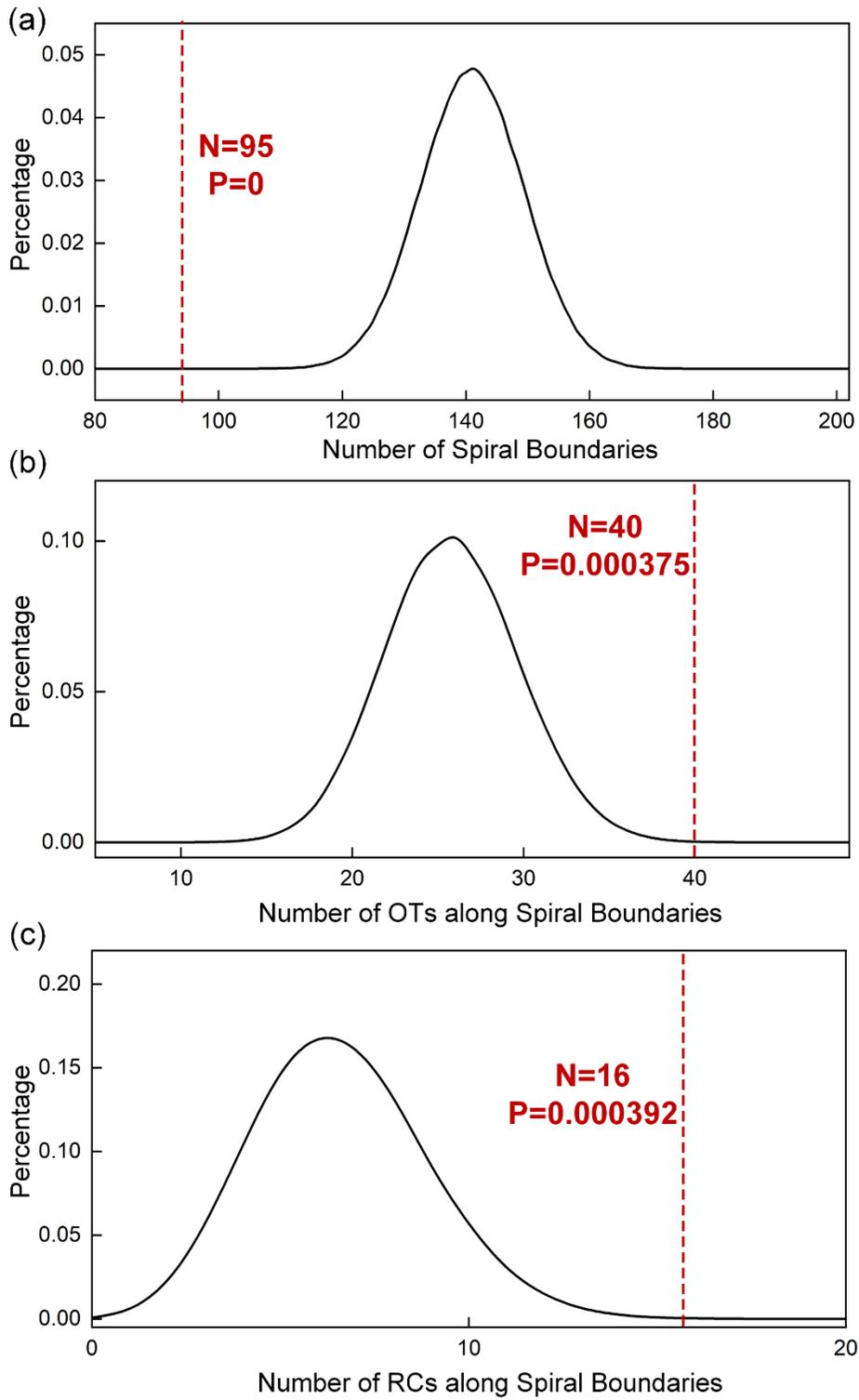

Figure S1.6 The distribution of probability density by modeling for the number of spiral boundaries, OTs and RCs.

2. The details of sample 2

The detailed images and data of sample 2 are shown from Figure S2.1-S2.6.

   a. The analysis of the non-random spiraling

As shown in Figure S2.3, there are 96 spiral boundaries and 152 non-spiral boundaries. Thus, the ratio of spiral boundaries is $96/(96 + 152) = 38.71\%$. The modeling based on the polygonal map shown in Figure S2.3 generates the probability density distribution shown in Figure S2.6(a). The event when the number of spiral boundaries is not higher than 96, which is the measured value, takes 0.0204%, indicating P≤0.000204.

   b. The analysis of OTs along spiral boundaries

As shown in Figure S2.4, there are 99 axons with discernible OTs along the spiral boundaries. Among them, there are 40 OTs along the spiral boundaries, indicated with purple circles. Thus, the ratio of OTs along spiral boundaries is $40/99 = 40.4\%$. The modeling based on the measurement of the ratio of spiral boundaries for each axons generates the probability density distribution shown in Figure S2.6(b). The event when the number of OTs along spiral boundaries is not lower than 40, which is the measured value, takes 0.63%, indicate $P \leq 0.0063$.

   c. The analysis of RCs along spiral boundaries

As shown in Figure S2.5, there are 51 groups of discernible RCs, while 17 of them are along the spiral boundaries, indicated with purple circles. Thus, the ratio of RCs along spiral boundaries is $17/51 = 33.33\%$. Meanwhile, the number of RCs along non-spiral boundaries is 23, leading to the ratio of RCs along non-spiral boundaries is $23/51 = 45.1\%$.

The total length of perimeter of all axons is 15030, while the total length of spiral boundaries and non-spiral boundaries are 2804 and 4544, respectively. Thus, the expected ratio of RCs along spiral and non-spiral boundaries should be $2804/15030 = 18.66\%$ and $4544/15030 = 30.23\%$ under the assumption of random distribution. This the measured value of ratio of RCs along spiral boundaries is significantly higher than the expected value ($33.33\%/18.66\% = 178.62\%$), while the measured value of ratio of RCs along non-spiral boundaries is also higher than the expected value ($45.1\%/30.23\% = 149.19\%$).

The modeling based on the expected ratio (18.66%) and the total number of RCs (51) generates the probability density distribution shown in Figure S2.6(c). The event when the number of RCs along spiral boundaries is not lower than 17, which is the measured value, takes 0.9%, indicate $P \leq 0.009$.

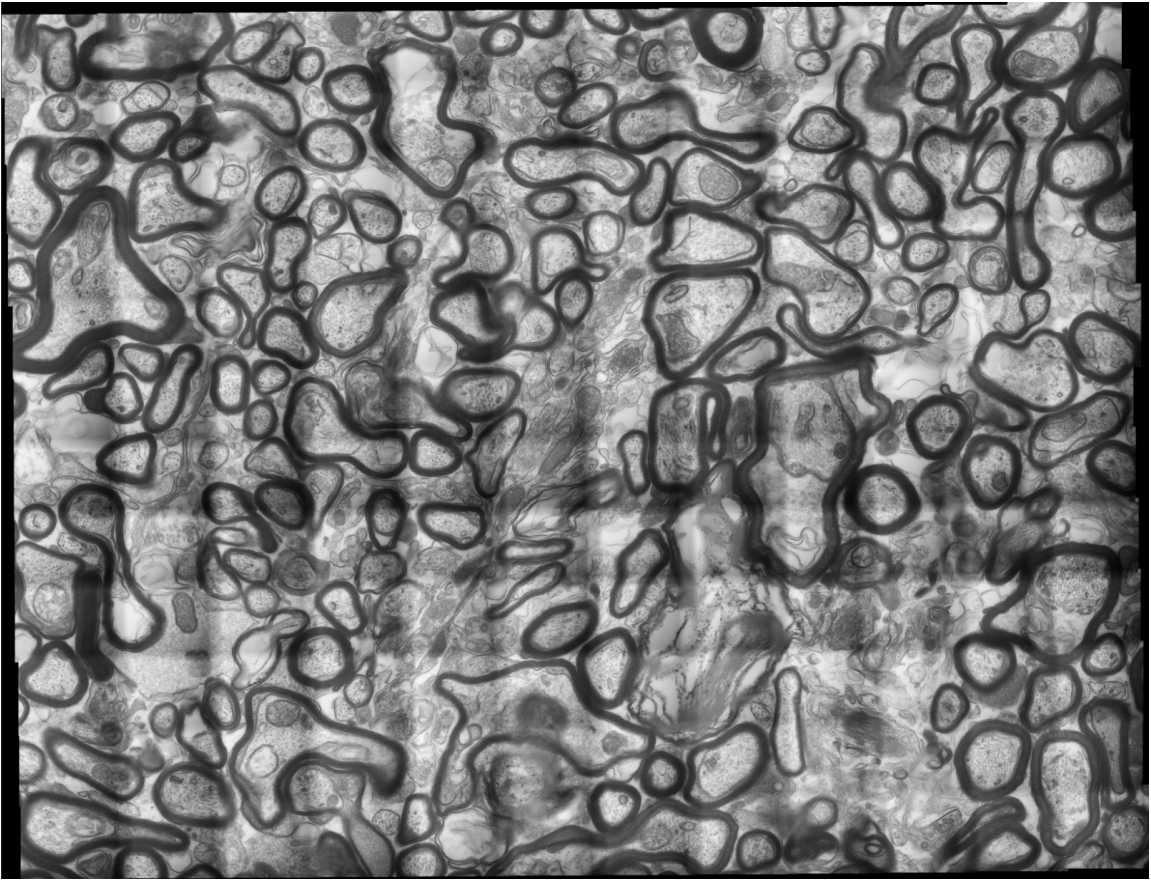

Figure S2.1 The ultra-large TEM image of sample 2 showing the optical nerve of mouse

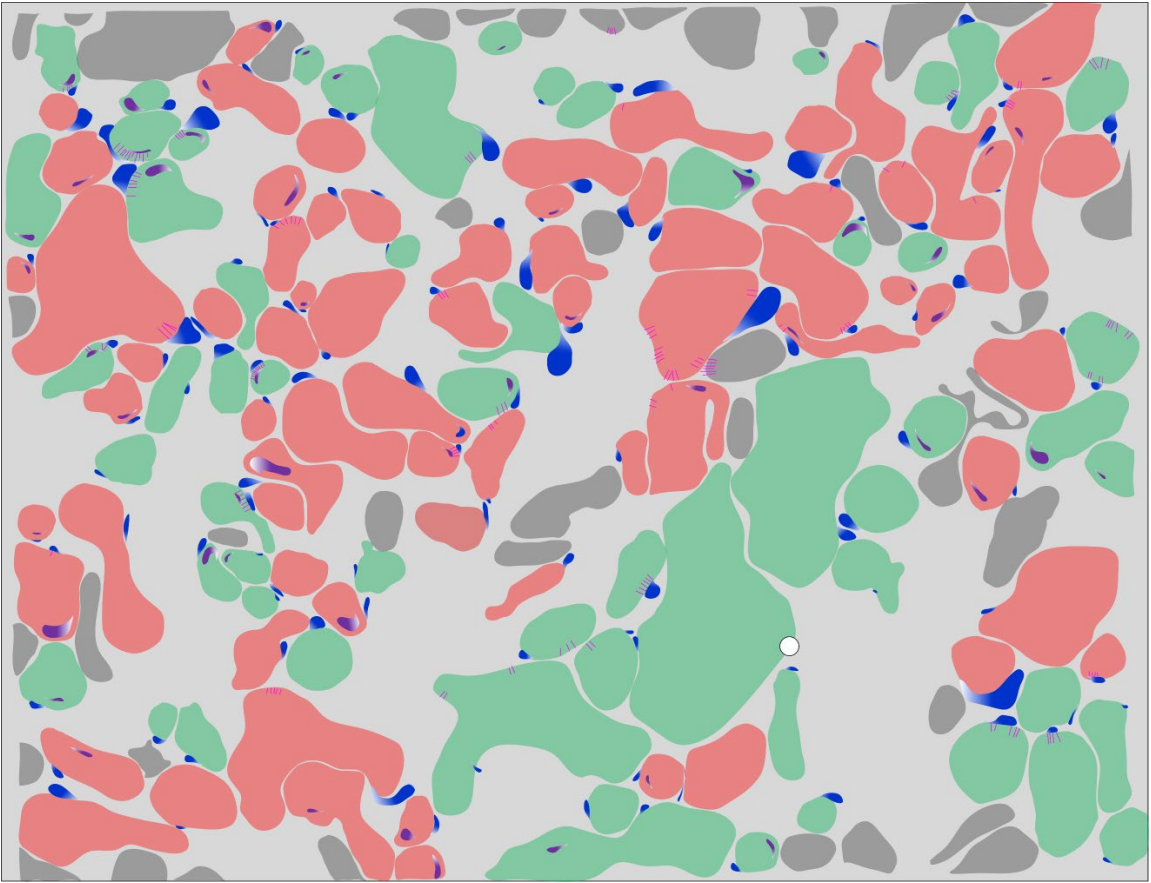

Figure S2.2 The spiral map of the TEM image of sample 2

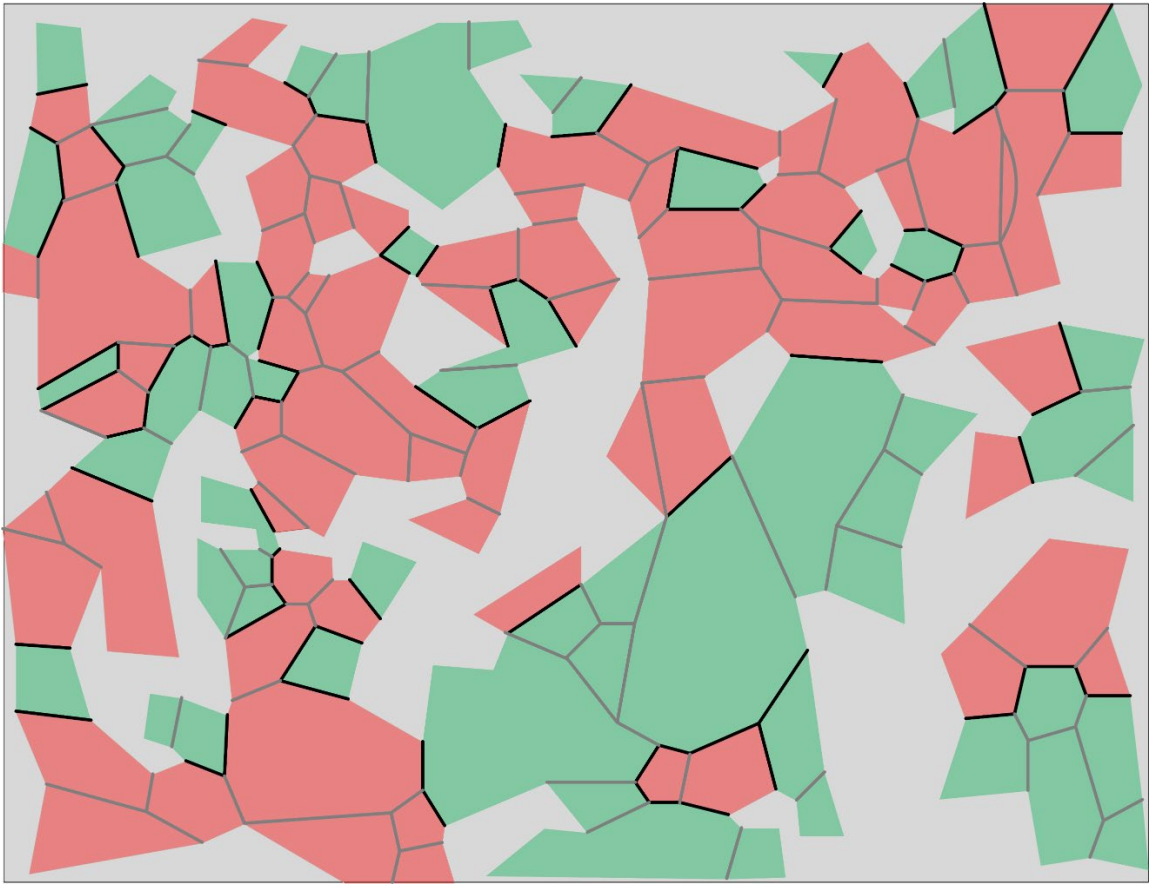

Figure S2.3 The polygonal map of sample 2

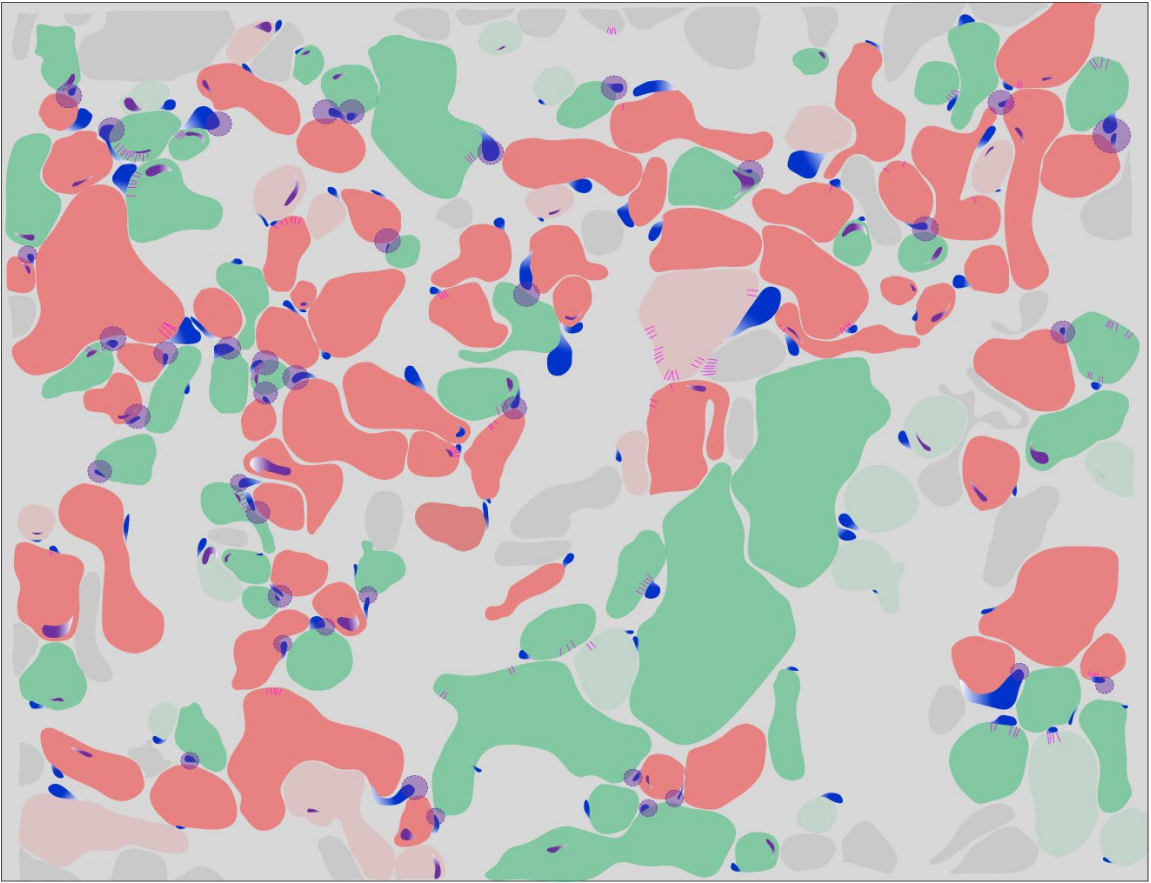

Figure S2.4 The axons along the spiral boundaries, with all 40 OTs at the spiral boundaries indicated with purple circles

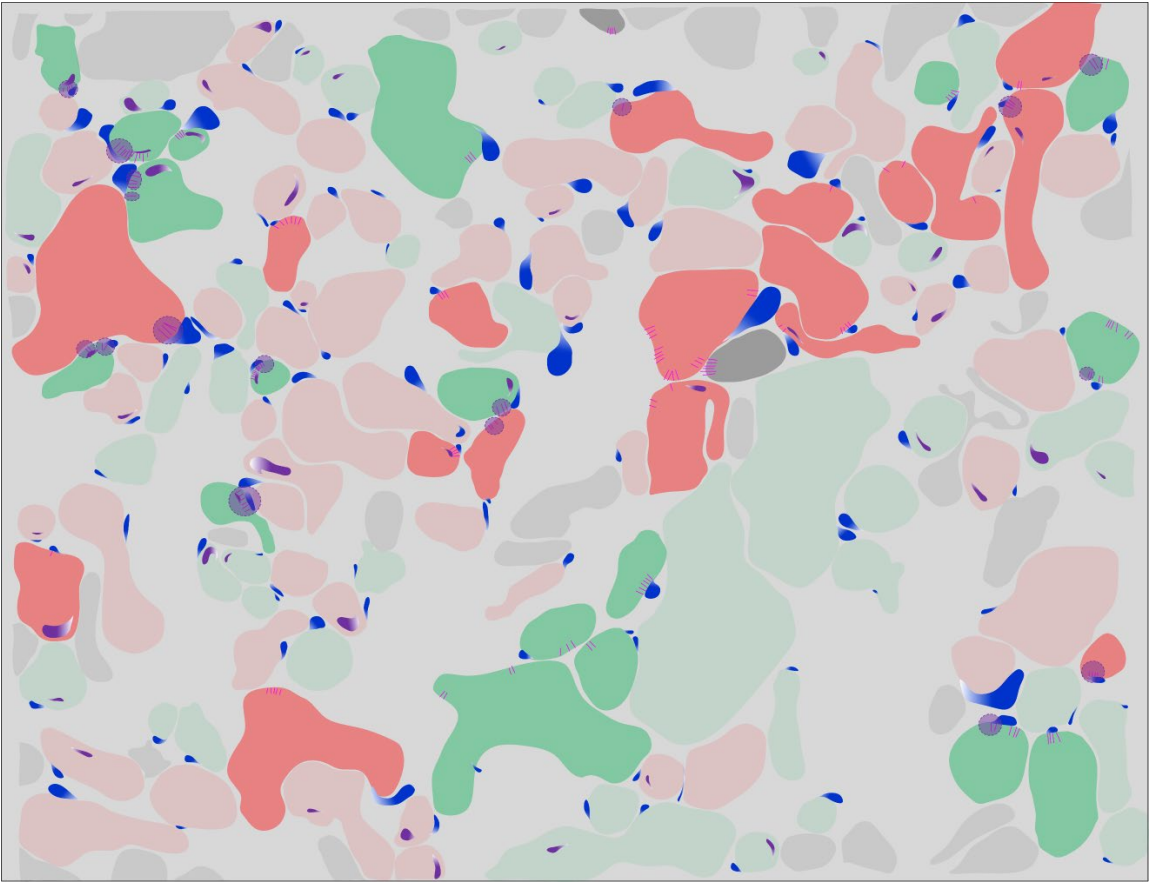

Figure S2.5 The axons with RCs, with all 17 RCs at the spiral boundaries indicated with purple circles

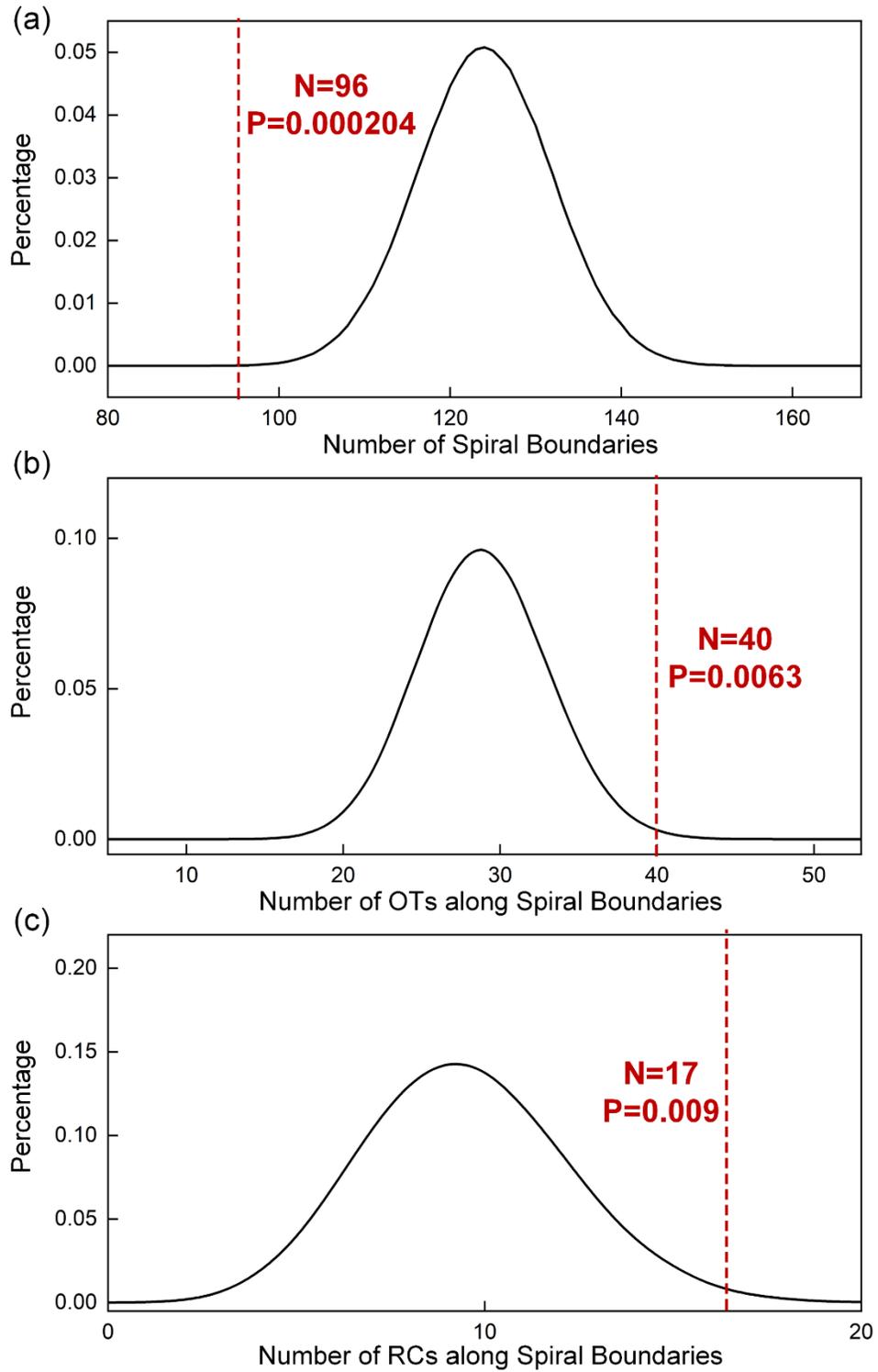

Figure S2.6 The distribution of probability density by modeling for the number of spiral boundaries, OTs and RCs.

3. The details of sample 3

The detailed images and data of sample 3 are shown from Figure S3.1-S3.6.

   a. The analysis of the non-random spiraling

As shown in Figure S3.3, there are 112 spiral boundaries and 156 non-spiral boundaries. Thus, the ratio of spiral boundaries is $112/(112 + 156) = 41.79\%$. The modeling based on the polygonal map shown in Figure S3.3 generates the probability density distribution shown in Figure S3.6(a). The event when the number of spiral boundaries is not higher than 112, which is the measured value, takes 0.44%, indicating $P \leq 0.0044$.

   b. The analysis of OTs along spiral boundaries

As shown in Figure S3.4, there are 94 axons with discernible OTs along the spiral boundaries. Among them, there are 40 OTs along the spiral boundaries, indicated with purple circles. Thus, the ratio of OTs along spiral boundaries is $40/94 = 42.55\%$. The modeling based on the measurement of the ratio of spiral boundaries for each axons generates the probability density distribution shown in Figure S3.6(b). The event when the number of OTs along spiral boundaries is not lower than 40, which is the measured value, takes 0.21%, indicate $P \leq 0.0021$.

   c. The analysis of RCs along spiral boundaries

As shown in Figure S3.5, there are 46 groups of discernible RCs, while 18 of them are along the spiral boundaries, indicated with purple circles. Thus, the ratio of RCs along spiral boundaries is $18/46 = 39.13\%$. Meanwhile, the number of RCs along non-spiral boundaries is 14, leading to the ratio of RCs along non-spiral boundaries is $14/46 = 30.43\%$.

The total length of perimeter of all axons is 14832, while the total length of spiral boundaries and non-spiral boundaries are 2900 and 4131, respectively. Thus, the expected ratio of RCs along spiral and non-spiral boundaries should be $2900/14832 = 19.55\%$ and $4131/14832 = 27.85\%$ under the assumption of random distribution. This the measured value of ratio of RCs along spiral boundaries is more than twice of the expected value ($39.13\%/19.55\% = 200.15\%$), while the measured value of ratio of RCs along non-spiral boundaries is also higher than the expected value ($30.43\%/27.85\% = 109.27\%$).

The modeling based on the expected ratio (19.55%) and the total number of RCs (46) generates the probability density distribution shown in Figure S3.6(c). The event when the number of RCs along spiral boundaries is not lower than 18, which is the measured value, takes 0.17%, indicate $P \leq 0.0017$.

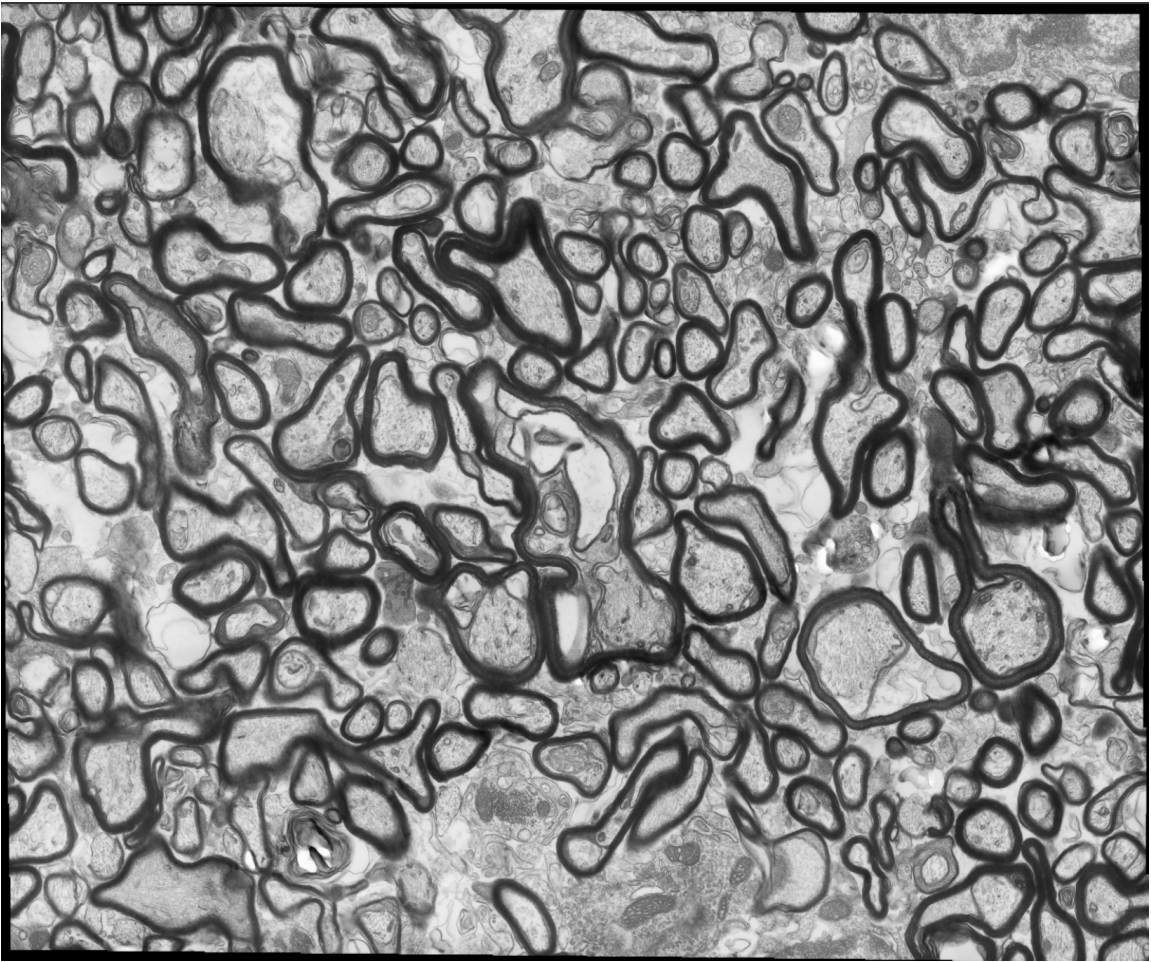

Figure S3.1 The ultra-large TEM image of sample 3 showing the optical nerve of mouse

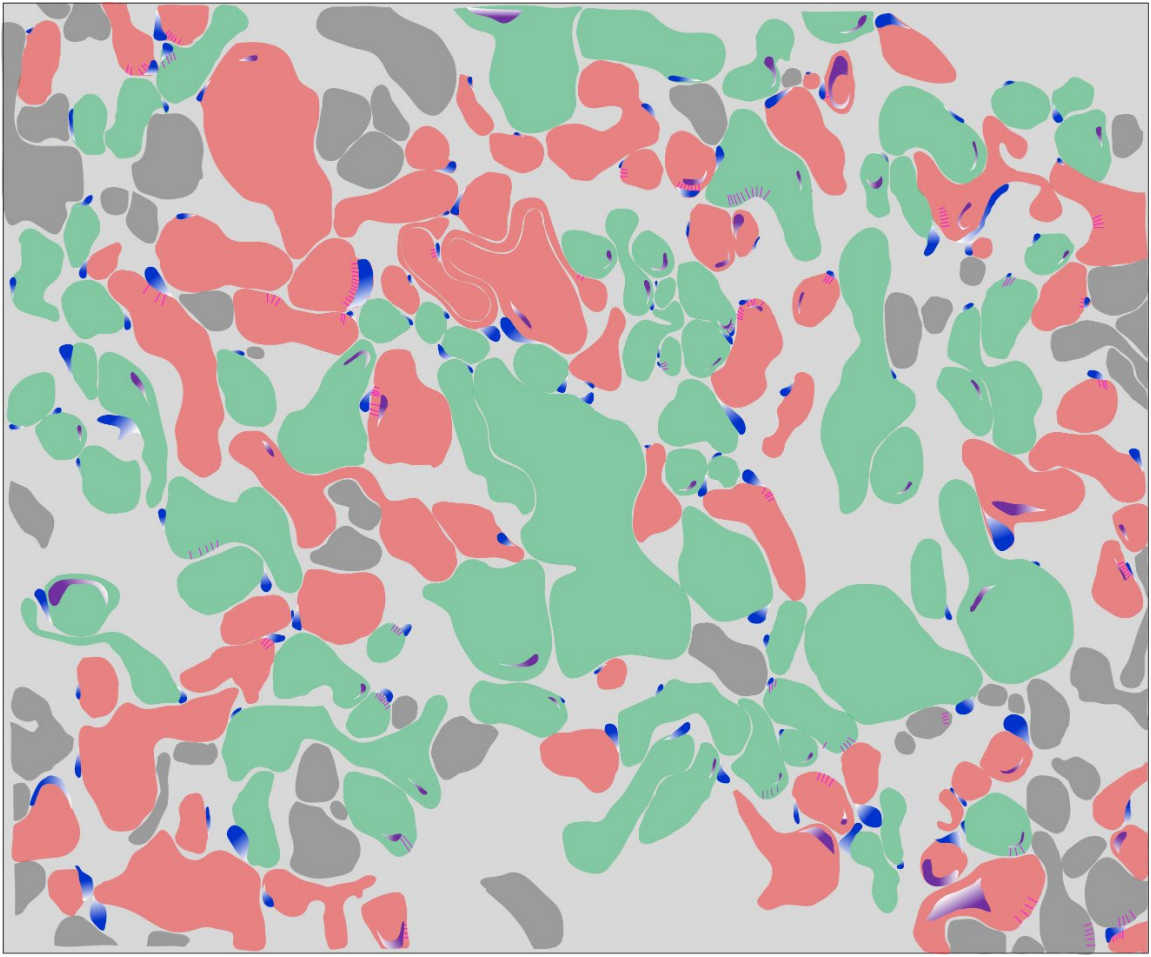

Figure S3.2 The spiral map of the TEM image of sample 3

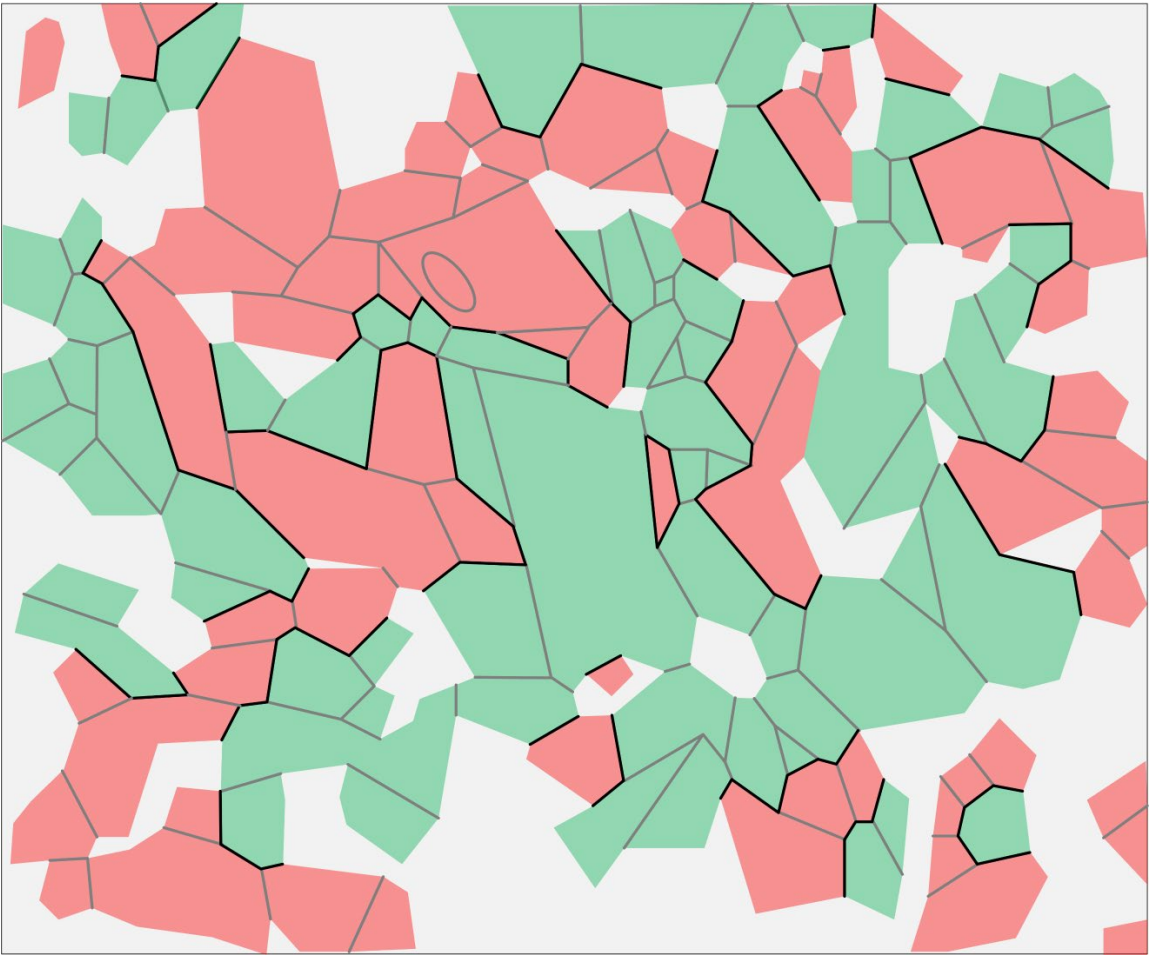

Figure S3.3 The polygonal map of sample 3

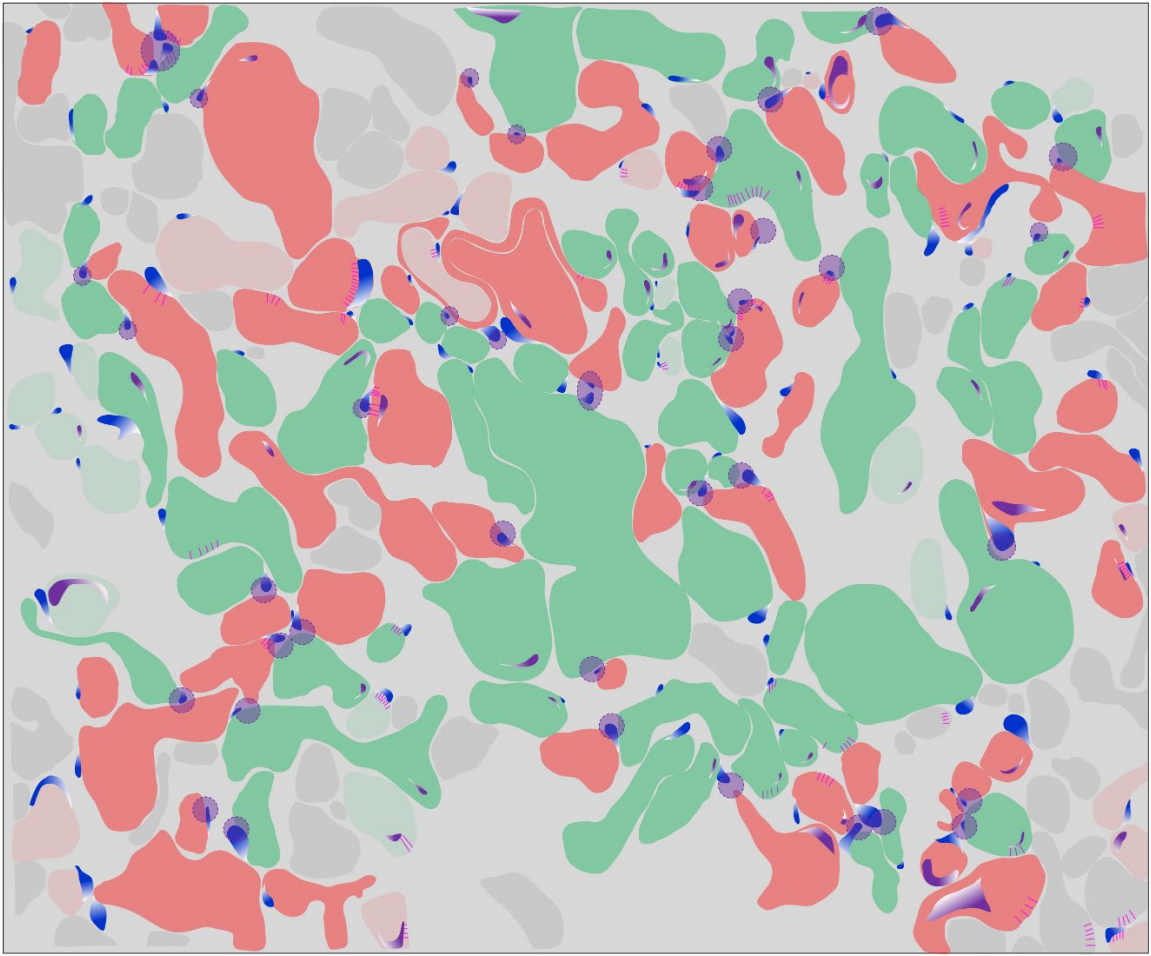

Figure S3.4 The axons along the spiral boundaries, with all 40 OTs at the spiral boundaries indicated with purple circles

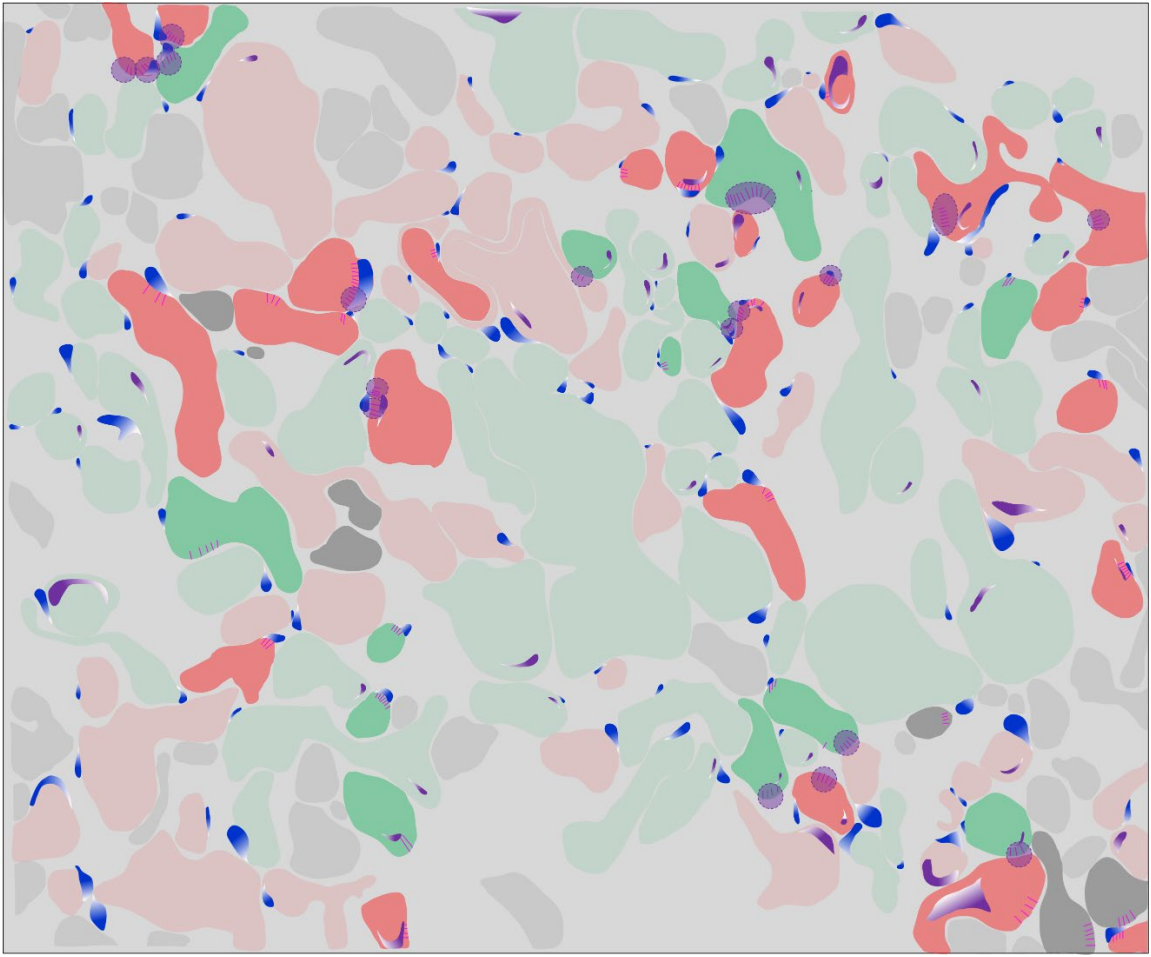

Figure S3.5 The axons with RCs, with all 18 RCs at the spiral boundaries indicated with purple circles

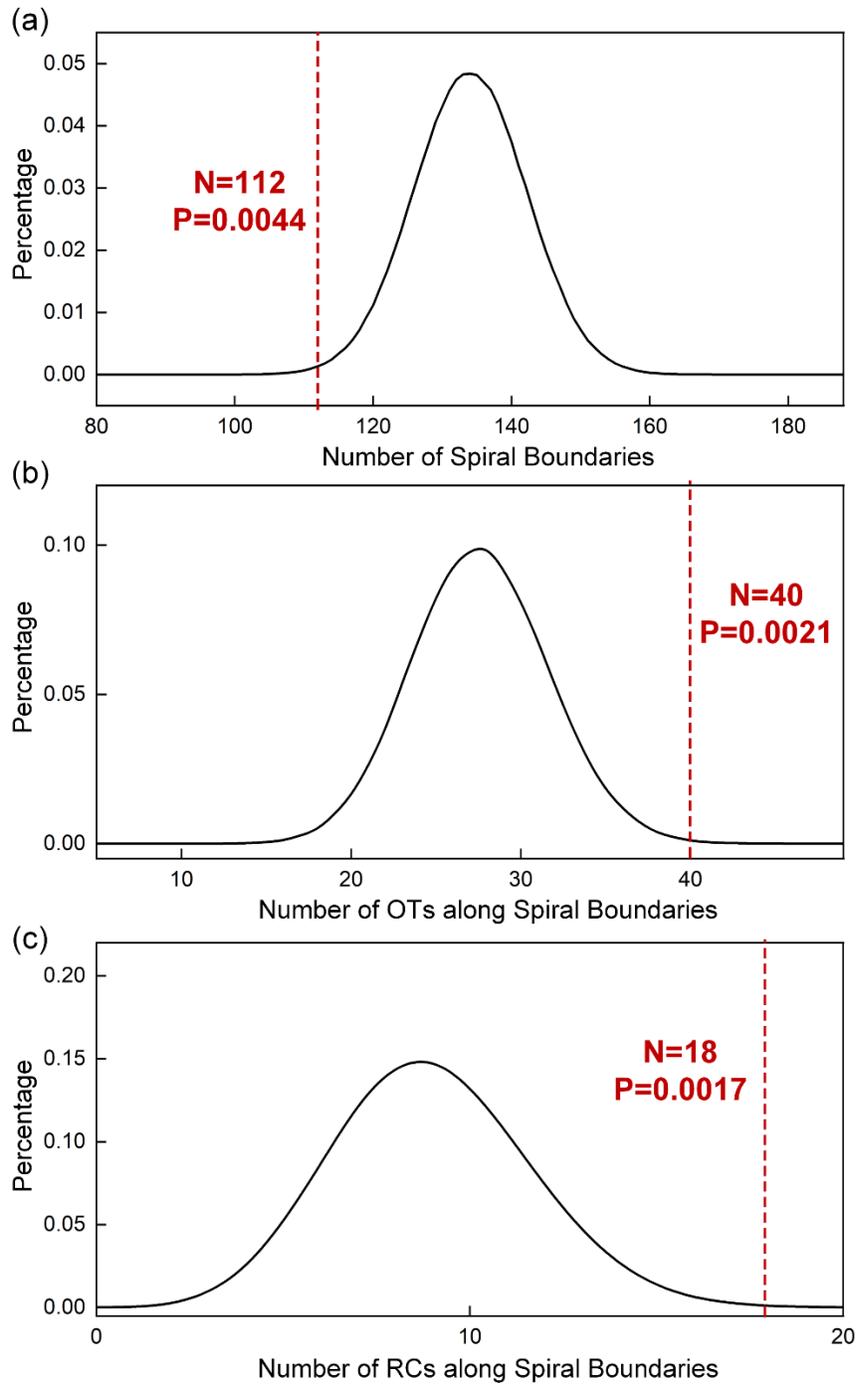

Figure S3.6 The distribution of probability density by modeling for the number of spiral boundaries, OTs and RCs.

**S2. Detailed consideration for the categorization of OTs and RCs to either spiral and non-spiral boundaries**

1. **Consideration of OTs**

We use sample 1 as an example for illustration. All the myelin sheaths along the spiral boundaries, along with their discernible OTs, are indicated as purple circles in Figure S4.1(a). Typical cases, as shown in Figure S4.1(b-d), frequently appear in the TEM images. In these instances, the OTs either remain aligned with the spiral boundary (Figure S4.1(b)), make contact at a point attributable to the spiral boundary (Figure S4.2(c)), or are positioned without direct contact but face an oppositely spiraling myelin sheath, which can also be attributed to the spiral boundary (Figure S4.1(d)). Following the same principle used to observe the same quadrant phenomenon[1], which involves measuring the position of the ITs, only the point of the leading edge of the OTs (red spots in Figure S4.1(b-d)) are considered for categorization as either spiral or non-spiral boundaries.

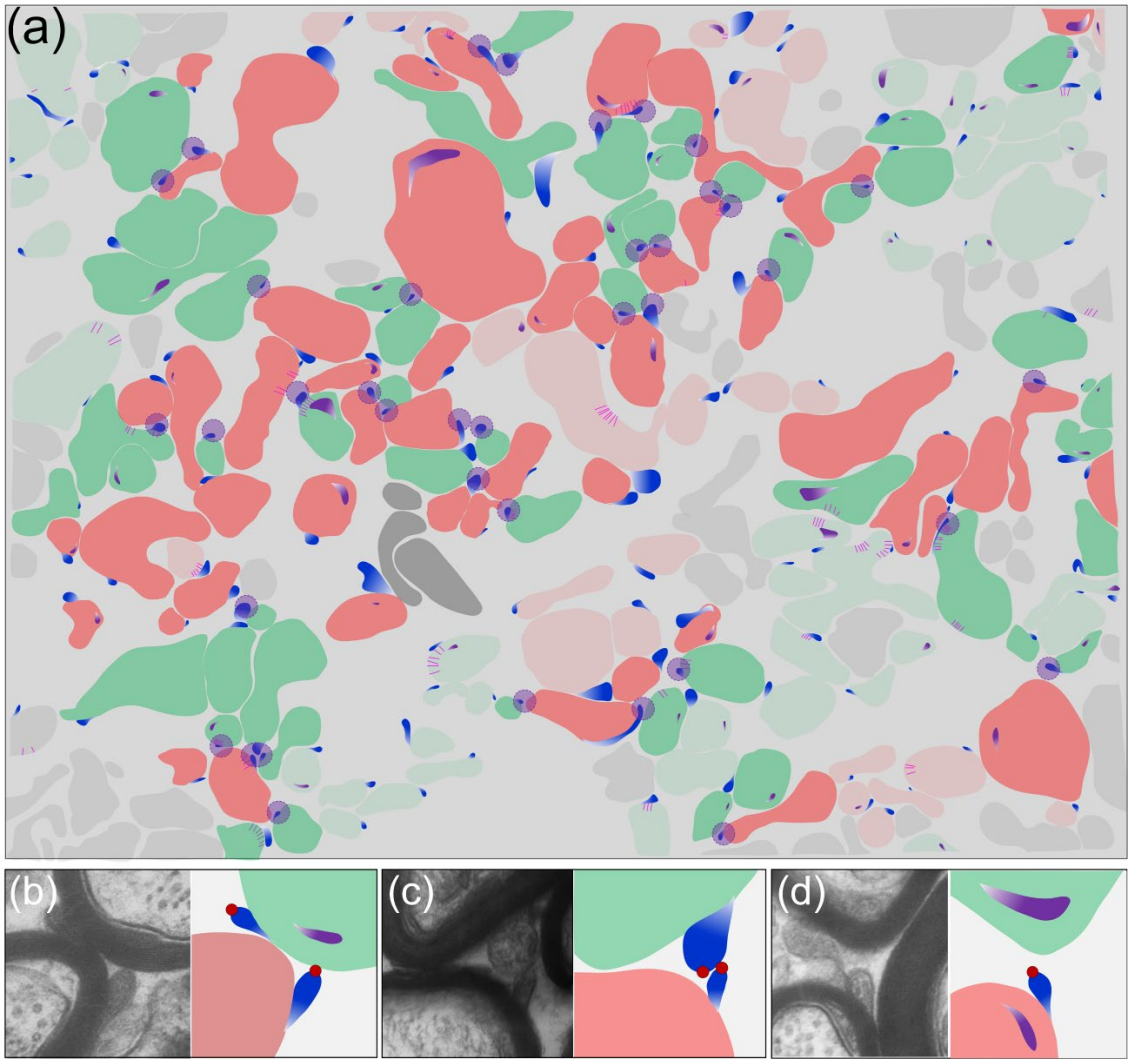

Figure S4.1 The OTs along spiral boundaries; (a) Details of the 40 OTs along spiral boundaries, all relevant axons along the spiral boundaries are with higher contrast; (b-d) Three possible situations that can be categorized to the group of OTs along spiral boundaries;

## 2. Consideration of RCs

We use Sample 1 for illustration, where all discernible RCs are indicated with purple circles (Figure S4.2(a)). Two types of RCs can influence the statistical results. Type 1, as shown in Figure S4.2(b), is entirely contained within either a spiral boundary or a non-spiral boundary, and will thus be categorized accordingly—either as an RC on a spiral boundary or a non-spiral boundary. Type 2 in Figure S4.2(c) represents a myelin sheath whose RCs are divided into two segments: one segment along a spiral boundary and the other along a non-spiral boundary. In this case, one count will be attributed to each group.

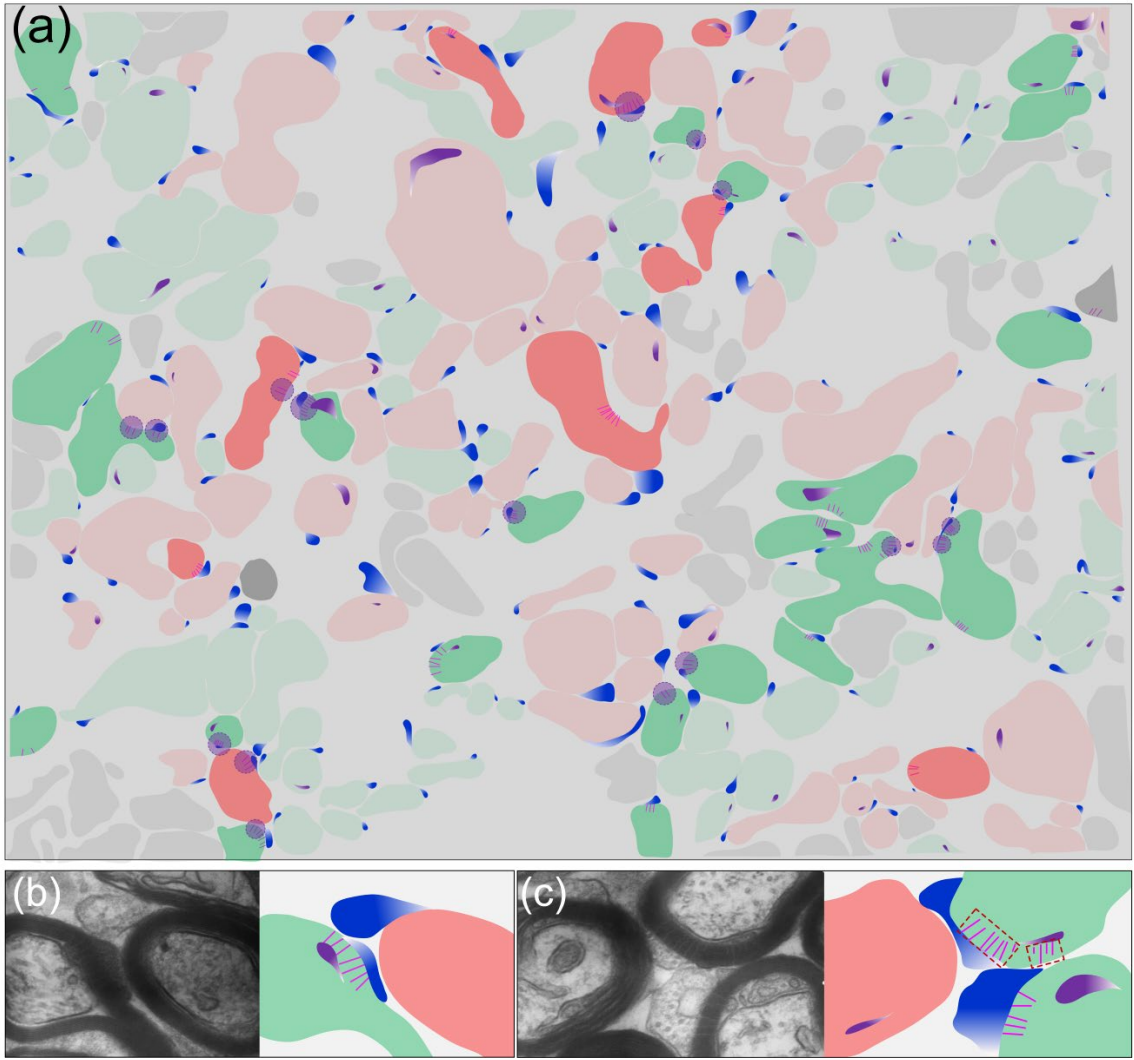

Figure S4.2 The RCs along the spial boundaries; (a) Details of 16 RCs along the spiral boundaries, all relevant axons with RCs are with higher contrast; (b-c) Two types of RCs that affect the categorization.

**Reference**

1. Peters, A., 1964. Further observations on the structure of myelin sheaths in the central nervous system. The Journal of cell biology, 20(2), pp.281-296.
2.